\documentclass[twocolumn,secnumarabic,amssymb, amsmath, nobibnotes, aps, prd]{revtex4-1}

\usepackage[dvipdfmx]{graphicx}  
\usepackage[dvipdfmx]{color}
\usepackage{dcolumn}   
\usepackage{bm}        
\usepackage{amssymb}   
\usepackage{here}
\usepackage{amsmath}
\usepackage{comment}
\usepackage{color}
\usepackage{braket}
\usepackage{mathrsfs}
\usepackage{amsmath}
\usepackage{ulem}
\usepackage{fancybox}
\usepackage{xspace}
\usepackage{amsfonts}
\usepackage{multirow}
\usepackage{amsthm}
\usepackage{setspace}
\usepackage[subrefformat=parens]{subcaption}
\usepackage{ulem}
\usepackage[version=3]{mhchem}

\begin{document}

\title{Three-dimensional topological magnon systems}%

\author{Hiroki Kondo, Yutaka Akagi, and Hosho Katsura}
\affiliation{Department of Physics, Graduate School of Science, The University of Tokyo, Hongo, Tokyo 113-0033, Japan}
\email[]{kondou@cams.phys.s.u-tokyo.ac.jp}

\begin{abstract}
We propose a concept of three-dimensional topological magnon systems which are the magnonic analog of topological insulators in three dimensions. We define a set of $\mathbb{Z}_2$ topological invariants that characterizes different topological phases and determines the presence or absence of surface Dirac cones. 
The validity of the classification scheme based on these $\mathbb{Z}_2$ invariants is demonstrated by considering three concrete examples.
One of them is a bosonic counterpart of the Fu-Kane-Mele model on a diamond lattice, which is found to exhibit three distinct phases analogous to strong topological, weak topological, and trivial insulator phases of the original fermionic model. 
We also discuss a possible realization of the thermal Hall effect of surface magnons in the presence of a  
magnetic field in proximity to a normal ferromagnet. The topological characterization in this paper can also be applied to other systems such as spin liquids or paramagnets described by Schwinger bosons.
\end{abstract}

\maketitle

\section{INTRODUCTION}

In recent years, there has been growing interest in studying topological phases of matter~\cite{Schnyder08,Kitaev09,Ryu10}. 
In particular, the discovery of quantum spin Hall insulators~\cite{Kane05a,Kane05b} attracted a great deal of attention.
The topological phases of quantum spin Hall insulators are characterized by the ${\mathbb Z}_2$ topological invariant~\cite{Fu06, Fukui07,  Fukui08, Qi08,  Wang10, Loring10, Fulga12, Sbierski14, Loring15, Katsura16, Akagi17, Katsura18}. 
The idea of quantum spin Hall insulators has been extended to three-dimensional (3D) topological insulators with protected gapless surface states~\cite{Fu07a,Moore07,Roy09,Guo09,Weeks10}. 
Previous studies on 3D topological insulators have revealed that they show richer properties than those in 2D.
For example, magnetoelectric properties of 3D topological insulators are described by axion electrodynamics~\cite{Qi08} and, as a consequence, an image magnetic monopole is induced by an electric charge brought near the surface~\cite{Qi09}.

As well as electrons, bosonic particles such as magnons~\cite{Fujimoto09, Katsura10, Matsumoto11a, Matsumoto14, Shindou13a, Shindou13b, Kim16, Onose10, Ideue12,Chisnell15, Hirschberger15, Han_Lee17, Murakami_Okamoto17,Kawano19}, photons~\cite{Onoda04, Hosten08, Raghu08, Haldane08, Wang09a, Ben-Abdallah16}, phonons~\cite{Strohm05, Sheng06, Inyushkin07, Kagan08, Wang09b, Zhang10, Qin12, Mori14, Huber16, Sugii17}, and triplons~\cite{Rumhanyi15,Joshi18}, have been shown to exhibit fascinating phenomena which 
originate from the topology of band structures.
Among them, magnons are of particular interest.
The research on topological magnon systems was triggered by the prediction of the thermal Hall effect of magnons~\cite{Katsura10}, which was soon observed in the pyrochlore ferromagnet Lu$_{2}$V$_{2}$O$_{7}$~\cite{Onose10}.
Other notable examples include the magnonic analog of spin Hall insulators~\cite{Zyuzin16,Nakata17}, and topological semimetal phases as well~\cite{Fransson16,Owerre17,Li16,Mook16,Su17}.
The topological properties of the magnon systems with a bulk band gap are characterized by the presence of gapless surface states.
In the previous study, we have defined the ${\mathbb Z}_2$ topological invariant that determines the presence or absence of magnonic helical edge states~\cite{Kondo19}. 
The ${\mathbb Z}_2$ topological invariant is defined based on the symmetry of the pseudo-time-reversal operator whose square is $-1$.
The studies of ${\mathbb Z}_2$ topological phases of magnons have so far been limited to 2D systems.  
However, as in the case of electrons, we expect that topologically protected surface states of magnons  in 3D exhibit novel phenomena arising from the coupling of their spin and orbital motion, which never appear in 2D.

In this paper, we propose three models of magnonic analog of 3D topological insulator, which we dub the 3D topological magnon systems. 
They are described by a bosonic Bogoliubov-de Gennes (BdG) Hamiltonian. Such systems are intrinsically non-Hermitian, as their Hamiltonians must be diagonalized by paraunitary matrices. 
According to the recent classification of  topological phases of non-Hermitian systems~\cite{Kawabata18,Zhou18}, our models belong to the symmetry class AII in 3D. 
Following our previous work~\cite{Kondo19}, we define a set of  topological invariants for 
these systems based on Kramers pairs in bosonic systems.
We discuss the correspondence of these topological invariants with the parity of the number of surface Dirac cones.
One of the three models is reminiscent of  the Fu-Kane-Mele model~\cite{Fu07a}.
Since this model has inversion symmetry, one can apply a bosonic counterpart of the simplified formula for the topological invariant~\cite{Fu07b}, allowing us to compute it analytically. 
At the end of this paper, we predict that the thermal Hall effect occurs in the presence of the surface magnetic field.

The rest of the paper is organized as follows.
In Sec.~II, we define a set of topological invariants that characterizes the topological phases of 3D topological magnon systems and discuss the correspondence between topological invariants and the surface states. 
In Sec.~III, we propose the first model of 3D topological magnon systems. The above correspondence is confirmed by calculating the band structure of the system with open boundary conditions. Next, in Sec.~IV, we discuss a possible realization of the thermal Hall effect  in the presence of a magnetic field on the surface in proximity to a normal ferromagnet. 
Concluding remarks are presented in Sec.~V.
In App.~A, we show the details of the interactions in the models which are proposed in Sec.~III.
In App.~B, the simplified expression of topological invariants for the systems with inversion symmetry is applied to this model.
In App.~C, we calculate analytically the energy spectrum of this model in a certain case.
In App.~D and E, we propose the second and third model of 3D topological magnon systems and confirm the correspondence between topological invariants and the surface states. 
In App.~F, we show the details of Dzyaloshinskii-Moriya (DM) interaction in the third model.

\section{DEFINITION OF TOPOLOGICAL INVARIANT}

We consider a system of non-interacting bosons in 3D lattices.
Assuming translational invariance, a generic quadratic Hamiltonian describing the system is given by
\begin{align}
\mathcal{H}=\frac{1}{2}\sum_{\bm{k}}[\bm{\beta}^{\dagger}(\bm{k})\bm{\beta}(-\bm{k})]H(\bm{k})
\left[
\begin{array}{cc}
\bm{\beta}(\bm{k})  \\
\bm{\beta}^{\dagger}(-\bm{k})  \\
\end{array} 
\right].
\label{eq:Ham}
\end{align}
Here, $\bm{\beta}^{\dagger}(\bm{k})=[\beta_{1}^{\dagger}(\bm{k}),\cdots,\beta_{\mathscr{N}}^{\dagger}(\bm{k})]$ denotes boson creation operators with momentum $\bm{k}=(k_{x},k_{y},k_{z})$. The subscript $\mathscr{N}$ is a number of internal degrees of freedom in a unit cell.
The $2\mathscr{N} \times 2\mathscr{N}$ Hermitian matrix $H(\bm{k})$ is a bosonic BdG Hamiltonian.

We consider the pseudo-time-reversal operator $\Theta'$ which is introduced in our previous work~\cite{Kondo19}.
This operator is written as $\Theta'=PK$, where $P$ and $K$ are a paraunitary matrix satisfying $P^{\dagger}\Sigma_{z}P=\Sigma_{z}$ and the complex conjugate operator, respectively.
Here, $\Sigma_{z}$ is defined as a tensor product $\Sigma_{z}:=\sigma_z \otimes  1_\mathscr{N}$ where $\sigma_{a}$ $(a=x,y,z)$ is the $a$-component of the Pauli matrix acting on the
particle-hole space and $1_\mathscr{N}$ is the $\mathscr{N} \times \mathscr{N}$ identity matrix. 
As shown in Ref~\cite{Kondo19}, a $2\mathscr{N}$-dimensional complex vector $\bm{\psi}$  and $\Theta'\bm{\psi}$ are orthogonal, i.e., $\langle \! \langle \bm{\psi},\Theta'\bm{\psi} \rangle \! \rangle=0$ 
under the scalar product defined by
\begin{align}
\langle \! \langle \bm{\phi},\bm{\psi} \rangle \! \rangle=\bm{\phi}^{\dagger}\Sigma_{z}\bm{\psi},
\end{align}
where, $\bm{\phi}^{\dagger}$ is the adjoint of the vector $\bm{\phi}$. 
Then the existence of bosonic Kramers degeneracy follows from the pseudo-time-reversal symmetry:
\begin{align}
\Sigma_{z}H(-\bm{k})\Theta'-\Theta'\Sigma_{z}H(\bm{k})=0. 
\label{eq:Commutation}
\end{align}

\begin{figure}[H]
\centering
  \includegraphics[width=8cm]{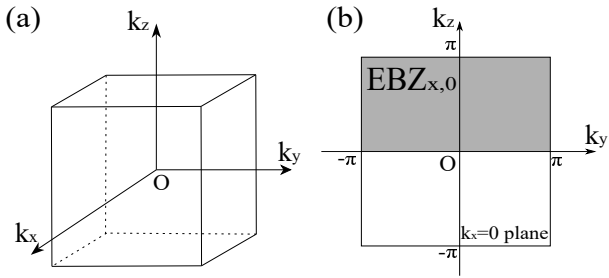}
\caption{(a) Brillouin zone of the simple cubic lattice: $k_{x}\in[-\pi,\pi],k_{y}\in[-\pi,\pi],k_{z}\in[-\pi,\pi]$.
(b) The region  corresponding to the effective Brillouin zone ${\rm EBZ}_{x,0}$: $k_{x}=0,k_{y}\in[-\pi,\pi],k_{z}\in[0,\pi]$.
}\label{fig:3DBZ_EBZ}
\end{figure}

Assuming the pseudo-time-reversal symmetry, we define the ${\mathbb Z}_{2}$ topological invariants for 3D magnon topological systems as follows:
\begin{align}
&\nu_{i,0}^{n\sigma} \!\! := \!\!\frac{1}{2\pi} \! \!\left[\! \oint_{\partial {\rm EBZ}_{i,0}} \!\!\!\!\!\!\!\!\!\!\!\!\!\!\!\!\! d\bm{k} \!\cdot \!\left[\bm{A}_{n\sigma}(\bm{k})\right]_{k_{i}=0} \!\!\! - \!\!\! \int_{{\rm EBZ}_{i,0}} \!\!\!\!\!\!\!\!\!\!\!\!\!\!\!\! dk_{j}dk_{k} \! \!\left[\Omega_{n\sigma}^{i}(\bm{k})\right]_{k_{i}=0} \!\right]\hspace{0mm}{\rm mod}\hspace{1mm}2, \nonumber\\
&\nu_{i,\pi}^{n\sigma} \!\! := \!\!\frac{1}{2\pi} \! \!\left[ \! \oint_{\partial {\rm EBZ}_{i,\pi}} \!\!\!\!\!\!\!\!\!\!\!\!\!\!\!\!\!\! d\bm{k} \!\cdot \!\left[\bm{A}_{n\sigma}(\bm{k})\right]_{k_{i}=\pi} \!\!\! - \!\!\! \int_{{\rm EBZ}_{i,\pi}} \!\!\!\!\!\!\!\!\!\!\!\!\!\!\!\! dk_{j}dk_{k} \! \!\left[\Omega_{n\sigma}^{i}(\bm{k})\right]_{k_{i}=\pi} \! \right]\hspace{0mm}{\rm mod}\hspace{1mm}2, \label{eq:topo_inv}
\end{align}
where $n$ is a band index and $i=x$, $y$, $z$.
Here, $j$ and $k$ represent two of $x$, $y$, $z$ which are different from $i$. 
The index $\sigma=\pm$ indicates the particle and hole spaces. The notation ${\rm EBZ}_{i,0}$ $({\rm EBZ}_{i,\pi})$ stands for the effective Brillouin zone in the $k_{i}=0$ $(k_{i}=\pi)$ plane, which is defined as $k_{i}=0$ $(k_{i}=\pi),k_{j}\in[-\pi,\pi],k_{k}\in[0,\pi]$.
Its boundary is denoted by $\partial {\rm EBZ}_{i,0}$ $(\partial {\rm EBZ}_{i,\pi})$.
3D Brillouin zone and one of the effective Brllouin zones, ${\rm EBZ}_{x,0}$, are shown in Fig.~\ref{fig:3DBZ_EBZ}.  Here we note that any 3D lattice can be deformed so that the Brillouin zone becomes a cube as shown in Fig.~\ref{fig:3DBZ_EBZ}(a) without changing the topology of the band structure.
Using an index $l (=1, 2)$ for Kramers doublet, $\bm{A}_{n\sigma}(\bm{k})$ and $\bm{\Omega}_{n\sigma}(\bm{k})$ are defined as
\begin{align}
&\bm{A}_{n\sigma}(\bm{k})=\sum_{l}\bm{A}_{nl\sigma}(\bm{k}),
\\
&\bm{\Omega}_{n\sigma}(\bm{k})=\sum_{l}\bm{\Omega}_{nl\sigma}(\bm{k}), 
\end{align}
where the Berry connection $\bm{A}_{n l \sigma}(\bm{k})$ and Berry curvature $\bm{\Omega}_{n l \sigma}(\bm{k})$ for bosons are given by 
\begin{align}
&\bm{A}_{nl\sigma}(\bm{k})
={\rm i} \left\langle \! \left\langle \bm{\Psi}_{nl\sigma}(\bm{k}),\nabla_{\bm{k}}\bm{\Psi}_{nl\sigma}(\bm{k})\right\rangle \! \right\rangle,
\\
&\bm{\Omega}_{nl\sigma}(\bm{k})=\nabla_{\bm{k}} \times \bm{A}_{nl\sigma}(\bm{k}).
\end{align} 
Here, $\bm{\Psi}_{nl\sigma}(\bm{k})$ is an eigenvector of $\Sigma_{z}H(\bm{k})$. 
As mentioned in Ref.~\cite{Kondo19}, the topological invariants of the particle and hole have the same value: $\nu_{i,0}^{n+}=\nu_{i,0}^{n-}, \nu_{i,\pi}^{n+}=\nu_{i,\pi}^{n-}$. 
Thus, it suffices to consider  the case $\sigma=+$. In the following, we drop $\sigma$ and simply write $\nu_{i,0}^{n+}$ and $\nu_{i,\pi}^{n+}$ as $\nu_{i,0}^{n}$ and $\nu_{i,\pi}^{n}$, respectively.

The number of the surface Dirac cones appearing in a bulk band gap can be understood in terms of the $\mathbb{Z}_2$ topological invariants. 
For a given bulk gap, oddness or evenness of the number of in-gap Dirac points at time-reversal-invariant momenta (TRIM) in ${\rm EBZ}_{i,0}$ $({\rm EBZ}_{i,\pi})$ is given by $\nu_{i,0}$ $(\nu_{i,\pi})$ with
\begin{align}
&\nu_{i,0}:=\sum_{n,(E_{n} ({\bm k}) \leq \epsilon_{0})}\nu_{i,0}^{n}\hspace{5mm}{\rm mod} \hspace{1mm}2,  \\
&\nu_{i,\pi}:=\sum_{n,(E_{n}({\bm k}) \leq \epsilon_{0})}\nu_{i,\pi}^{n}\hspace{5mm}{\rm mod} \hspace{1mm}2,
\end{align}
where $E_n ({\bm k})$ is the energy of the $n$th band and $\epsilon_0$ is a fictitious Fermi energy that can be arbitrary as long as it lies in the gap. 
Since $\nu_{i,0}+\nu_{i,\pi}$ corresponds to the parity of the total number of the surface Dirac points, one has
\begin{align}
\nu_{x,0}+\nu_{x,\pi}=\nu_{y,0}+\nu_{y,\pi}=\nu_{z,0}+\nu_{z,\pi} \hspace{5mm}{\rm mod} \hspace{1mm}2.
\end{align}
Due to this equation, only four of the six topological invariants are independent.
Thus the topological phases of the system are completely characterized by the set of four topological invariants: $(\nu_{0};\nu_{x},\nu_{y},\nu_{z})$, where  
\begin{align}
&\nu_{0}=\nu_{x,0}+\nu_{x,\pi} \hspace{5mm}{\rm mod} \hspace{1mm}2, \\
&\nu_{i}=\nu_{i,\pi}\hspace{5mm}(i=x,y,z).
\end{align}
Following the discussion in Ref.~\cite{Fu07a}, for $\nu_{0}=1$, there exist odd number of surface Dirac cones in total. Such a topological phase is robust against disorder respecting pseudo-time-reversal symmetry and corresponds to the strong topological phase of electronic topological insulators. If the four topological invariants are all zero, the system is in the trivial phase.
When $\nu_{0}=0$ but at least one of $\nu_{i}$ $(i=x,y,z)$ is nonzero, there exist even number of surface Dirac cones in total. 
However, this phase is not robust even against perturbations preserving pseudo-time-reversal symmetry, because even number of surface Dirac cones can be paired up and removed.

\section{MODELS}
\subsection{Hamiltonian}

Now we propose a model Hamiltonian of 3D topological magnon systems.
We assume that the bosons from up and down spins have the same number of internal degrees of freedom.
For such systems, the set of boson creation operators $\bm{\beta}^{\dagger}(\bm{k})$ in
Eq.~(\ref{eq:Ham}) takes the form: 
\begin{align}
\bm{\beta}^{\dagger}(\bm{k})=[\bm{b}_{\uparrow}^{\dagger}(\bm{k}),\bm{b}_{\downarrow}^{\dagger}(\bm{k})].
\end{align}
Here, the creation operators of boson from up spins $\bm{b}_{\uparrow}^{\dagger}(\bm{k})$ and those from down spins $\bm{b}_{\downarrow}^{\dagger}(\bm{k})$ are given by
\begin{align}
&\bm{b}_{\uparrow}^{\dagger}(\bm{k})=[b_{\uparrow,1}^{\dagger}(\bm{k}),\cdots,b_{\uparrow,N}^{\dagger}(\bm{k})],   \nonumber \\
&\bm{b}_{\downarrow}^{\dagger}(\bm{k})=[b_{\downarrow,1}^{\dagger}(\bm{k}),\cdots,b_{\downarrow,N}^{\dagger}(\bm{k})],
\label{eq:opvec}
\end{align}
where $N$ is the number of the internal degrees of freedom. The operator $b_{s,i}^{\dagger}(\bm{k})$ creates a boson with spin $s$ ($s=\uparrow, \downarrow$) at site $i$. 
In this setup, we take the pseudo-time-reversal operator to be of the form
\begin{align}
\Theta'=(\sigma_{z} \otimes {\rm i}\sigma_{y} \otimes 1_{N})K.
\end{align}

\begin{figure}[H]
\centering
  \includegraphics[width=6cm]{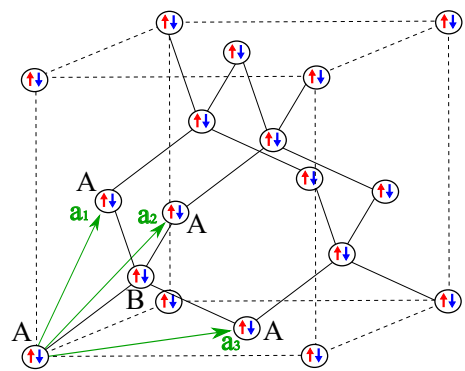}
\caption{Diamond lattice system having two spins at each site. Two sublattices are denoted by $A$ and $B$. The vectors $\bm{a}_{i}$ $(i=1,2,3)$ are the lattice prime vectors. 
}\label{fig:cell_di}
\end{figure}

Here we introduce our first example, which is a system in which two spins with opposite directions are localized at each site of the diamond lattice. 
The system is depicted in Fig.~\ref{fig:cell_di}.
The Hamiltonian of the system is written as
\begin{align}
H=H_{\rm DM}+H_{J'}+H_{J}+H_{\rm XY}+H_{\rm \Gamma}+H_{\kappa} .\label{eq:Ham_di}
\end{align}
Here, $H_{\rm DM}$ is the DM interaction between next-nearest-neighbor spins which point  in the same direction.
The second term $H_{J'}$ is the antiferromagnetic interaction between two spins on the same lattice site. 
The third term $H_{J}$ is the bond-dependent ferromagnetic interaction between nearest-neighbor spins aligned in the same direction.
The magnon band structure opens a gap due to the bond dependence of this interaction.
The terms $H_{\rm XY}$ and $H_{\rm \Gamma}$ are the anisotropic XY and ${\rm \Gamma}$ interactions between next-nearest-neighbor spins pointing in opposite directions.
The last term $H_{\kappa}$ is the single ion anisotropy.
The interactions above realize the Fu-Kane-Mele model ~\cite{Fu07a} in magnon systems.
Ferromagnetic Heisenberg interaction between nearest neighbor sites ($H_{J}$) plays the role of the hopping term $t \sum_{\langle i,j\rangle}c_{i}^{\dagger}c_{j}$ in Eq.~(4) in Ref.~\cite{Fu07a}.
The DM interaction ($H_{{\rm DM}}$), the Gamma interaction ($H_{{\rm \Gamma}}$), and the Heisenberg interaction with XY anisotropy ($H_{{\rm XY}}$) are considered to mimic the spin-orbit interaction $i(8\lambda_{SO}/a^{2})\sum_{\langle\!\langle i,j\rangle\!\rangle}c_{i}^{\dagger}\bm{s}\cdot(\bm{d}_{ij}^{1}\times \bm{d}_{ij}^{2})c_{j}$ in Eq.~(4) in Ref.~\cite{Fu07a}.
To sum up, the correspondence between the magnetic and the fermionic Hamiltonians is as follows:
\begin{align}
&H_{J} \leftrightarrow t \sum_{\langle i,j\rangle}c_{i}^{\dagger}c_{j} , \\
&H_{{\rm DM}} \leftrightarrow i(8\lambda_{SO}/a^{2})\sum_{\langle\!\langle i,j\rangle\!\rangle}c_{i}^{\dagger}s_{z}(\bm{d}_{ij}^{1}\times \bm{d}_{ij}^{2})_{z}c_{j} , \\
&H_{{\rm \Gamma}} \leftrightarrow i(8\lambda_{SO}/a^{2})\sum_{\langle\!\langle i,j\rangle\!\rangle}c_{i}^{\dagger}s_{x}(\bm{d}_{ij}^{1}\times \bm{d}_{ij}^{2})_{x}c_{j} , \\
&H_{{\rm XY}} \leftrightarrow i(8\lambda_{SO}/a^{2})\sum_{\langle\!\langle i,j\rangle\!\rangle}c_{i}^{\dagger}s_{y}(\bm{d}_{ij}^{1}\times \bm{d}_{ij}^{2})_{y}c_{j} .
\end{align}
As shown in Section III C,  the interaction $H_{J'}$ does not change the topology of the band structure.
The easy axis anisotropy $H_{\kappa}$ is needed to make the spin configuration in Fig.~\ref{fig:cell_di} stable.
The details of these interactions are shown in Appendix A.

\noindent

By applying the Holstein-Primakoff and Fourier transformation, we can write down the Hamiltonian in the same form as Eq.~(\ref{eq:Ham}) with $N=2$. The matrix $H(\bm{k})$ here is proportional to the length of spins $S$.
In the following, we take $S=1$ for simplicity. The explicit expression for the matrix $H(\bm{k})$ is written as follows:
\begin{align}
H(\bm{k})=
\left(
\begin{array}{cc}
h(\bm{k}) &\Delta^{\dagger}(\bm{k}) \\
\Delta(\bm{k}) &h^{*}(-\bm{k}) \\
\end{array} 
\right),
\label{eq:H_FKM}
\end{align}
where $h(\bm{k})$ and $\Delta(\bm{k})$ are $4\times 4$ matrices defined as
\begin{align}
&h(\bm{k})=d_{0}1_{4}+\sum_{a=1}^{5}d_{a}(\bm{k})\Gamma^{a},  \label{eq:h}\\
&\Delta(\bm{k})=J'\sigma_{x}\otimes 1_{2},
\end{align}
with
\begin{align}
\Gamma^{(1,2,3,4,5)}=(1_{2}\otimes\sigma_{x} ,1_{2}\otimes \sigma_{y},\sigma_{x}\otimes\sigma_{z} ,\sigma_{y}\otimes\sigma_{z} ,\sigma_{z}\otimes \sigma_{z}).\nonumber 
\end{align}
These $\Gamma$ matrices satisfy the standard anticommutation relations: $\{\Gamma^{(i)},\Gamma^{(j)}  \} =2 \delta_{ij} 1_{4} $.
The six components in Eq.~(\ref{eq:h}) are defined as 
\begin{align}
&d_{0}=J_{0}+J_{1}+J_{2}+J_{3}+J'+2\kappa ,\nonumber \\
&d_{1}(\bm{k})=-J_{0}-J_{1}\cos{(k_{1})}-J_{2}\cos{(k_{2})}-J_{3}\cos{(k_{3})} ,\nonumber \\
&d_{2}(\bm{k})=-J_{1}\sin{(k_{1})}-J_{2}\sin{(k_{2})}-J_{3}\sin{(k_{3})} , \nonumber \\
&d_{3}(\bm{k})=-\sqrt{2}\Gamma  [\sin{(k_{2})}-\sin{(k_{3})}-\sin{(k_{21})}+\sin{(k_{31})}]  ,\nonumber \\
&d_{4}(\bm{k})=-\sqrt{2}J_{-} [\sin{(k_{3})}-\sin{(k_{1})}-\sin{(k_{32})}+\sin{(k_{12})}]  ,\nonumber \\
&d_{5}(\bm{k})=-\sqrt{2}D [\sin{(k_{1})}-\sin{(k_{2})}-\sin{(k_{13})}+\sin{(k_{23})}] ,\nonumber 
\end{align}
where $k_{i}=\bm{k}\cdot \bm{a}_{i}$ and $k_{ij}=k_{i}-k_{j}$ and all the parameters are real. The vectors $\bm{a}_{i}$ $(i=1,2,3)$ are the three lattice prime vectors of the diamond lattice.
We note that the Hamiltonian~(\ref{eq:H_FKM}) satisfies the pseudo-time-reversal symmetry~(\ref{eq:Commutation}). 

\subsection{$\mathbb{Z}_2$ topological invariants}

Let us now compute the $\mathbb{Z}_2$ topological invariants. Thanks to the inversion symmetry of the system, one can compute them analytically by using a simplified formula which can be thought of as the bosonic counterpart of the formula derived in Ref.~\cite{Fu07b} (see Appendix B for details). The results are summarized in Table~\ref{table:index_di}.
We have checked that the topological invariants obtained are the same as those obtained by evaluating Eq.~(\ref{eq:topo_inv}) numerically.
Table~\ref{table:index_di} suggests that the system is in the strong topological phase, i.e., odd number of Dirac cones exist between the top and bottom bands in the particle space (the correspondence of topological invariants to the number of Dirac cones is confirmed later). 
The bulk band structure with the same parameters as those of Table~\ref{table:index_di} is shown in Fig.~\ref{fig:bandPBC_di}(a). Since the system has both the pseudo-time-reversal symmetry and inversion symmetry, the energy spectrum is doubly degenerate over the whole Brillouin zone.
We note that nontrivial topological phases can be realized when at least two of $J_{-},D$, and $\Gamma$ are nonzero \cite{Energy_analytical}.

\begin{table}[H]
\caption{The $\mathbb{Z}_2$ topological invariants of the model. 
The parameters are chosen to be 
$J_{0}=1.4,J_{1}=J_{2}=J_{3}=J'=1.0,J_{-}=D=\Gamma=0.3,\kappa=1.5$.
The index $n=1,2$ denotes the top and bottom band in the particle space, respectively.
}
\begin{center}
{\tabcolsep=4mm
  \begin{tabular}{cccccccc}
    \hline
    $n$  & 1 & 2 \\ \hline
    $\nu_{x,0}^{n}$  & 0 & 0 \\ \hline 
    $\nu_{x,\pi}^{n}$  & 1 & 1 \\ \hline 
    $\nu_{y,0}^{n}$  & 0 & 0 \\ \hline    
    $\nu_{y,\pi}^{n}$  & 1 & 1 \\ \hline 
    $\nu_{z,0}^{n}$  & 0 & 0 \\ \hline 
    $\nu_{z,\pi}^{n}$  & 1 &1 \\ \hline 
    $(\nu_{0}^{n};\nu_{x}^{n},\nu_{y}^{n},\nu_{z}^{n})$  & (1;1,1,1) &(1;1,1,1) \\ \hline 
  \end{tabular}
}
\end{center}
\label{table:index_di}
\end{table}
\noindent

\subsection{Phase diagram and magnon band structure}

We now construct a phase diagram of the diamond lattice system by computing the simplified formula of topological invariants analytically.
The resulting phase diagram is shown in Fig.~\ref{fig:bandPBC_di}(b) 
(see Appendix B for a detailed derivation).
By changing the coupling constants, three phases: strong topological, weak topological, and trivial phases are all realized.
The band structures for a slab with (100) face of the system which is deformed to a cubic lattice are shown in Fig.~\ref{fig:bandOBC_di}. 
As we expect from the general arguments of the relation between the topological invariants and the number of the surface Dirac points.
In weak topological phases $(0;111)$ and $(0;100)$ (Fig.~\ref{fig:bandOBC_di}(a) and (b), respectively), there are even number (2 and 0, respectively) of surface Dirac cones.
On the other hand, in strong topological phases $(1;111)$ and $(1;100)$ (Fig.~\ref{fig:bandOBC_di}(c) and (d), respectively), there are odd number (1 and 3, respectively) of surface Dirac cones. 
On general grounds, it is expected that the surface states arising in the strong topological phase exhibit magnon spin-momentum locking~\cite{Okuma17} even in the collinear magnet due to the terms $H_{\rm \Gamma}$ and $H_{\rm XY}$ which break spin conservation.

Let us now show that the magnonic system with nonzero $J'$ is in the topological phase corresponding to that of the Fu-Kane-Mele model~\cite{Fu07a}, whose Hamiltonian is essentially the same as Eq. (\ref{eq:H_FKM}) with $J'=0$.
Numerical results suggest that the band gap of $H({\bm k})$ closes only at TRIM, in which case this magnon system is adiabatically connected to the system with $J'=0$ without closing a gap. This can be seen as follows. 
We first note that the band gap at TRIM $\bm{k}=\bm{\Gamma}$ can be written as
\begin{align}
\Delta E=\sqrt{(d_{0}+|d_{1}(\bm{\Gamma})|)^{2}-J'^{2}}-\sqrt{(d_{0}-|d_{1}(\bm{\Gamma})|)^{2}-J'^{2}}, \nonumber 
\end{align}
(see Appendix B for a detailed derivation). 
Clearly, $\Delta E=0$ if $d_{1}(\bm{\Gamma})=0$ and $\Delta E > 0$ otherwise. 
Now suppose that the gap opens at $J' \neq 0$. This implies $d_{1}(\bm{\Gamma}) \neq 0$ at all TRIM. Then one finds that the gap does not close even if one takes $J'\to 0$. This proves the desired result.  

\begin{figure}[H]
\centering
  \includegraphics[width=8.8cm]{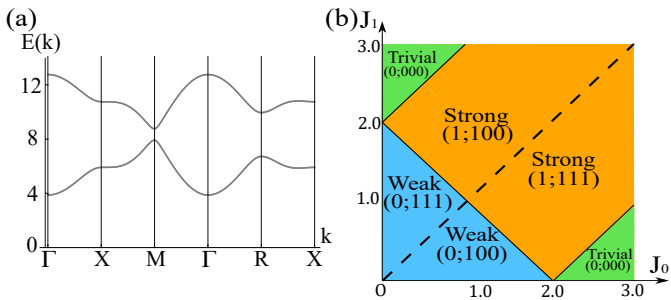}
\caption{(a) The bulk band structure of the system. Taking  $\bm{a}_{1}=(1,0,0)$, $\bm{a}_{2}=(0,1,0)$, and $\bm{a}_{3}=(0,0,1)$, we here deform the diamond lattice into 
an equivalent cubic lattice. Parameters are chosen to be $J_{0}=1.4$, $J_{1}=J_{2}=J_{3}=J'=1.0$, $J_{-}=D=\Gamma=0.3$, $\kappa=1.5$. The symmetry points are $\Gamma=(0,0,0)$, $X=(\pi,0,0)$, $M=(\pi,0,\pi)$, and $R=(\pi,\pi,\pi)$.
(b) The phase diagram of the system as a function of $J_{0}$ and $J_{1}$. 
In each phase, the corresponding topological indices are indicated by $(\nu_{0}^{2};\nu_{x}^{2},\nu_{y}^{2},\nu_{z}^{2})$. 
The other parameters are chosen to be $J_{2}=J_{3}=J'=1.0$, $J_{-}=D=\Gamma=0.3$, $\kappa=1.5$.
The dashed line indicates the phase boundary between two phases with different weak indices.
}\label{fig:bandPBC_di}
\end{figure}
\noindent

\begin{figure}[H]
\centering
  \includegraphics[width=9cm]{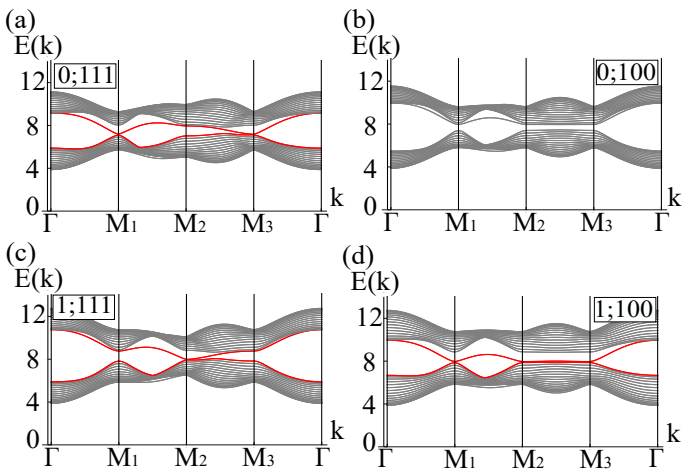}
\caption{Band structure for a slab with (100) face for the four phases in Fig.~\ref{fig:bandPBC_di}(b). 
The symmetry points are $\Gamma=(0,0),M_{1}=(\pi,0),M_{2}=(\pi,\pi)$, and $M_{3}=(0,\pi)$.
The topologically protected surface states are shown in red. Coupling constants $(J_{0},J_{1})$ are chosen to be (a) $(J_{0},J_{1})=(0.6, 1.0)$,  (b) $(J_{0},J_{1})=(1.0, 0.8)$,  (c) $(J_{0},J_{1})=(1.4, 1.0)$, and  (d) $(J_{0},J_{1})=(1.0, 1.4)$, respectively. The other parameters are the same as those in Fig.~\ref{fig:bandPBC_di}(b): $J_{2}=J_{3}=J'=1.0$, $J_{-}=D=\Gamma=0.3$, $\kappa=1.5$.
}\label{fig:bandOBC_di}
\end{figure}
\noindent

\subsection{Other models}

So far we have focused on the model on the diamond lattice. However, the physics of the 3D topological magnon systems is not limited to this lattice. To illustrate this, we have constructed two other examples: one is defined on the perovskite lattice and the other on the the pyrochlore lattice. In both cases, the validity of the relation between the set of topological invariants and the number of the surface Dirac points has been confirmed (see Appendix D and E for details).

\begin{center}
\bf{IV. THERMAL HALL EFFECT}
\end{center}

\begin{figure}[H]
\centering
  \includegraphics[width=5cm]{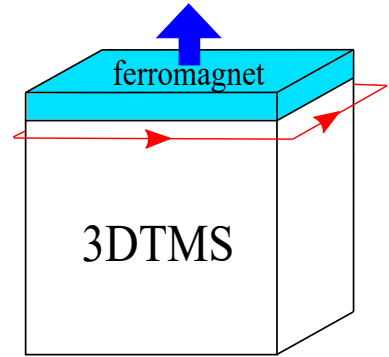}
\caption{The heterostructure of a ferromagnet and a 3D topological magnon system (3DTMS). The blue arrow indicates the direction of the magnetization of the ferromagnet. The chiral edge state shown by red arrows is responsible for the thermal Hall effect.
}\label{fig:3DTMS}
\end{figure}

We now turn to discuss the physical implications of surface Dirac cones in the strong topological phase. 
For electronic systems, previous studies have shown that the surface Dirac dispersions in a 3D topological insulator can be gapped out by applying a magnetic field on the surface~\cite{Sinitsyn07,Lu10}.
In general, such states have nonzero Berry curvature, leading to the surface quantum Hall effect. 
Here we show that the magnonic counterpart of this phenomenon occurs in our system. 
To this end, we consider a heterostructure of a ferromagnet and a 3D topological magnon system shown in Fig.~\ref{fig:3DTMS}.
To be concrete, we take the diamond lattice system as a 3D topological magnon system. 
We assume that the magnetic field emanating from the ferromagnet is weak enough not to affect the magnetic order in the substrate.

\noindent

\begin{figure}[H]
\centering
  \includegraphics[width=9cm]{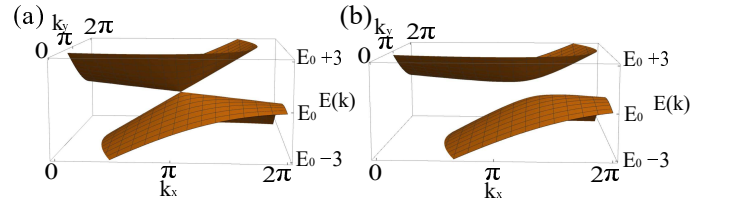}
\caption{Band structure of the $(001)$ surface state of the diamond lattice system (a) without and (b) with the surface magnetic field $BS=1.0$, where $E_{0}=8.34$ is the band touching energy.
Parameters are chosen to be  $J_{0}S=1.4,J_{1}S=J_{2}S=J_{3}S=J'S=1.0,J_{-}S=DS=\Gamma S=0.3, \kappa S=1.5$. 
The band structure is plotted in the range of $0\leq k_{x},k_{y}\leq2\pi$. 
}\label{fig:diamond_surface}
\end{figure}

\begin{figure}[H]
\centering
  \includegraphics[width=9cm]{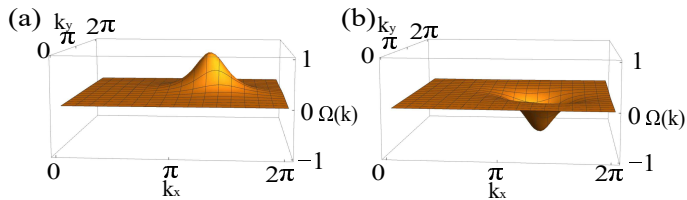}
\caption{Berry curvatures of  (a) the top and (b) the bottom bands of the $(001)$ surface state. Parameters are chosen to be  $J_{0}S=1.4,J_{1}S=J_{2}S=J_{3}S=J'S=1.0,J_{-}S=DS=\Gamma S=0.3, \kappa S=1.5, BS=1.0$, which correspond to those in Fig.~\ref{fig:diamond_surface}(b). We note that the two Berry curvatures are the same in magnitude but opposite in sign.
}\label{fig:diamond_surface_Berry}
\end{figure}

First, we derive the effective Hamiltonian of the surface of the system.
Let us consider the projection onto the subspace of two (particle) bands above and below the Dirac point we focus on.
Denoting the indices of the two bands as $n$ and $n+1$, we define the set of the wave functions of these bands as 
\begin{align}
&\ket{\bm{\Psi}(\bm{k})} \nonumber \\
&:=\left[\ket{\bm{\Psi}_{n,1,+}(\bm{k})}, \ket{\bm{\Psi}_{n,2,+}(\bm{k})}, \ket{\bm{\Psi}_{n+1,1,+}(\bm{k})}, \ket{\bm{\Psi}_{n+1,2,+}(\bm{k})}  \right].
\end{align}
The Hamiltonian projected onto this subspace is a $4\times 4$ matrix and written as
\begin{align}
H_{n}(\bm{k})=\bra{\bm{\Psi}(\bm{k})}H(\bm{k}) \ket{\bm{\Psi}(\bm{k})}.
\end{align}
Expanding the Hamiltonian $H_{n}(\bm{k})$ to the second order in $k_{z}$ and the first order in $(k_{x}-\pi)$ and ($k_{y}-\pi)$,
we denote the terms depending on $(k_{x}-\pi)$ or $(k_{y}-\pi)$ as $H_{xy}(k_{x},k_{y})$. The other terms are denoted by $H_{z}(k_{z})$.
We write the energy of the center of the Dirac cone as $E_{0}$.
The state $\bm{\psi}(z)$ which is localized on the $(001)$ surface is obtained as the eigenvector of the Hamiltonian $H_{z}(-{\rm i} \partial_{z})$ with the eigenvalue $E_{0}$:
\begin{align}
H_{z}(-{\rm i} \partial_{z})\bm{\psi}(z)=E_{0}\bm{\psi}(z).
\label{eq:surf}
\end{align}
Here we take $\bm{\psi}(z)$ as
\begin{align}
\bm{\psi}(z)=\bm{\psi}_{0}e^{\lambda z}.
\end{align}
The parameter $\lambda$ is obtained by solving
\begin{align}
{\rm Det}\left( H_{z}(-{\rm i}\lambda)-E_{0}1_{4} \right)=0.
\end{align}
This is an eighth order equation in $\lambda$ and has eight solutions. 
Four of them are positive and the other are negative. 
The negative solutions correspond to states localized on the $(00\bar{1})$ surface.
Here we focus on the positive solutions corresponding to states localized on the $(001)$ surface.
Due to the time-reversal symmetry, each two of the eight solutions are the same.
Thus, we write two positive solutions as $\lambda_{1}$ and $\lambda_{2}$ and define the corresponding two constant vectors as $\bm{\psi}_{1}$ and $\bm{\psi}_{2}$.
The pseudo-time-reversal operator in the subspace of $n$th and $(n+1)$th bands is defined as $\Theta=({\rm i}\sigma_{y} \otimes 1_{2})K$.
The vectors $\Theta\bm{\psi}_{1}$ and $\Theta\bm{\psi}_{2}$ are also the solutions with $\lambda_{1}$ and $\lambda_{2}$, respectively.
Then the wave function which satisfies Eq.~(\ref{eq:surf})  is generally written as
\begin{align}
\bm{\psi}(z)=\left(\alpha_{1}\bm{\psi}_{1}+\beta_{1}\Theta\bm{\psi}_{1}\right)e^{\lambda_{1} z}+\left(\alpha_{2}\bm{\psi}_{2}+\beta_{2}\Theta\bm{\psi}_{2}\right)e^{\lambda_{2} z}.
\end{align}
Since $\lambda_{1,2}>0$, this satisfies the condition $\bm{\psi}(-\infty)=0$. 
Using the other boundary condition that the wave function vanishes on the surface $z=0$, i.e. $\bm{\psi}(0)=0$, we obtain four simultaneous equations for the coefficient $\alpha_{1},\beta_{1},\alpha_{2}$ and $\beta_{2}$:
\begin{align}
\left(\bm{\psi}_{1},\Theta\bm{\psi}_{1},\bm{\psi}_{2},\Theta\bm{\psi}_{2} \right)
\left(
\begin{array}{cccc}
\alpha_{1}  \\
\beta_{1}  \\
\alpha_{2}  \\
\beta_{2}
\end{array}
\right)
=0.
\end{align}
Since the determinant of the matrix $\left(\bm{\psi}_{1},\Theta\bm{\psi}_{1},\bm{\psi}_{2},\Theta\bm{\psi}_{2} \right)$ is zero,  the above equation has a nontrivial solution.
The overall factor of the four coefficients $\alpha_{1},\beta_{1},\alpha_{2}$ and $\beta_{2}$ is determined by the normalization condition:
\begin{align}
\int_{-\infty}^{0}dz|\bm{\psi}(z)|^{2}=1.
\end{align}
From the coefficients obtained, we have the effective surface Hamiltonian
\begin{align}
&H_{\rm eff}(k_{x},k_{y}) \nonumber \\
&=\left(
\begin{array}{cc}
\bra{\bm{\psi}}H_{xy}(k_{x},k_{y})\ket{\bm{\psi}}+E_{0} & \bra{\bm{\psi}}H_{xy}(k_{x},k_{y})\ket{\Theta\bm{\psi}} \\
\bra{\Theta\bm{\psi}}H_{xy}(k_{x},k_{y})\ket{\bm{\psi}} &\bra{\Theta\bm{\psi}}H_{xy}(k_{x},k_{y})\ket{\Theta\bm{\psi}}+E_{0} 
\end{array}
\right),
\end{align}
where the matrix element $\bra{\bm{\psi}}H_{xy}(k_{x},k_{y})\ket{\bm{\psi}}$ is defined as
\begin{align}
\bra{\bm{\psi}}H_{xy}(k_{x},k_{y})\ket{\bm{\psi}}=\int_{-\infty}^{0}dz\bm{\psi}^{\dagger}(z)H_{xy}(k_{x},k_{y})\bm{\psi}(z).
\end{align}
The other three matrix elements are defined similarly.
By applying the magnetic field $\bm{B}=B\bm{e}_{z}$, the additional term $-BS\sigma_{z}$ appears in the Hamiltonian.
The band structure of the surface is obtained by diagonalizing the Hamiltonian $H_{\rm eff}(k_{x},k_{y})$.
The Berry curvature of the system is defined as $\Omega_{n}^{z}(k_{x},k_{y})=2{\rm Im}\left[(\partial_{k_{x}}\bm{\psi}_{n}^{\dagger}(k_{x},k_{y}))(\partial_{k_{y}}\bm{\psi}_{n}(k_{x},k_{y}))\right]$.
Figure~\ref{fig:diamond_surface} shows the band structure of the surface Hamiltonian without and with the surface magnetic field. We can see that the surface state can be gapped out by applying the surface magnetic field.
Figure~\ref{fig:diamond_surface_Berry} shows the Berry curvature of the top and bottom bands under magnetic field, respectively.
Since the surface state has nonzero Berry curvature, the thermal Hall coefficient calculated in Ref \cite{Matsumoto11b}:
\begin{align}
\kappa^{xy}=\frac{k_{B}^{2}T}{\hbar V}\sum_{n,k_{x},k_{y}}c_{2}(\rho(E_{n}(k_{x},k_{y})))\Omega_{n}^{z}(k_{x},k_{y})
,\label{eq:thermal_Hall_coefficient}
\end{align}
is expected to be nonzero. Here, $k_{B},\hbar, V,$ and $T$ are the Boltzmann constant, the Planck constant, the volume of the system, and the average temperature of the system, respectively.
The function $c_{2}(\rho)$ is defined as $c_{2}(\rho)=(1+\rho)\left(\log\frac{1+\rho}{\rho}\right)^{2}-\left(\log\rho\right)^{2}-2{\rm Li}_{2}(-\rho)$, where ${\rm Li}_{2}(-\rho)$ and $\rho(E_{n}(k_{x},k_{y}))$ are the dilogarithm function and the Bose distribution function, respectively.

\noindent
\noindent

\begin{center}
\bf{V. SUMMARY}
\end{center}

In this paper, we proposed three models of 3D topological magnon systems that host topologically protected surface Dirac cones. 
The models have pseudo-time-reversal symmetry which allows us to define the set of $\mathbb{Z}_2$ topological invariants that can distinguish different phases of these systems.
For a model on the diamond lattice, we computed the topological invariants analytically by exploiting the inversion symmetry of the system. The analysis of these topological invariants revealed that the model exhibits strong topological, weak topological, and trivial phases, in which the number of surface Dirac points is odd, even, and zero, respectively. 

We also studied the topological invariants in the other two models numerically and found the same correspondence. 
As a physical consequence of the presence of surface Dirac dispersions, we discussed the thermal Hall effect in the composite system consisting of a ferromagnet and a 3D topological magnon system. 
In this setup, the Zeeman field at the interface gaps out the surface Dirac dispersions, leading to 
the nonvanishing Berry curvature, and hence a nonzero thermal Hall response. 
Finally, we remark that the present formulation of the $\mathbb{Z}_2$ topological invariants can be applied to other bosonic systems such as phonons and photons, as long as they respect pseudo-time-reversal symmetry. In this respect, it would be interesting to study bosonic excitations in spin liquids or paramagnets, which are entirely different from the magnon excitations. We expect that bosonic $\mathbb{Z}_2$ topological phases in such systems can be identified by combining our approach with the Schwinger-boson mean-field theory~\cite{Sachdev92, Wang06, Liu19}.

\begin{center}
\bf{Acknowledgements}
\end{center}

This work was supported by JSPS KAKENHI Grants No. JP18H04478, No. JP18K03445, JP17K14352, and JP18H04220.
H. Kondo was supported by the  JSPS through Program for Leading Graduate Schools (ALPS).

\begin{center}
\bf{Appendix A: Details of the interactions}
\end{center}

\begin{figure}[H]
\centering
  \includegraphics[width=7cm]{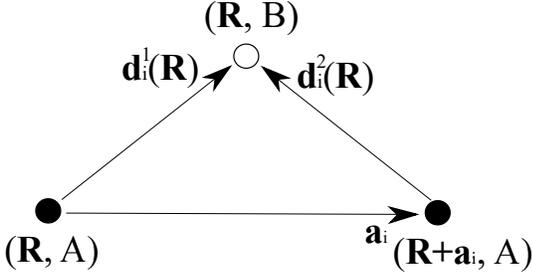}
\caption{Two nearest neighbor bond vectors $\bm{d}^{1}_{i}(\bm{R})$ and $\bm{d}^{2}_{i}(\bm{R})$.
The site $(\bm{R},B)$ is the common nearest neighbor site between two next nearest sites $(\bm{R},A)$ and $(\bm{R}+\bm{a}_{i},A)$. The vectors  $\bm{d}^{1}_{i}(\bm{R})$ and $\bm{d}^{2}_{i}(\bm{R})$ point to the site $(\bm{R},B)$ from $(\bm{R},A)$ and $(\bm{R}+\bm{a}_{i},A)$, respectively.
}\label{fig:DMvector}
\end{figure}

In this part, we describe the details of the interactions of the first model (\ref{eq:Ham_di}).
They are written as follows:
\begin{widetext}
\begin{align}
&H_{\rm DM}
=\sum_{\bm{R},s={\rm u,d}}
D_{1}^{z}(S_{s}^{x}(\bm{R},A)S_{s}^{y}(\bm{R}+\bm{a}_{1},A)-S_{s}^{y}(\bm{R},A)S_{s}^{x}(\bm{R}+\bm{a}_{1},A)) \nonumber \\
&\hspace{20mm}+D_{2}^{z}(S_{s}^{x}(\bm{R},A)S_{s}^{y}(\bm{R}+\bm{a}_{2},A)-S_{s}^{y}(\bm{R},A)S_{s}^{x}(\bm{R}+\bm{a}_{2},A)) \nonumber \\
&\hspace{20mm}+D_{3}^{z}(S_{s}^{x}(\bm{R},A)S_{s}^{y}(\bm{R}+\bm{a}_{3},A)-S_{s}^{y}(\bm{R},A)S_{s}^{x}(\bm{R}+\bm{a}_{3},A)) \nonumber \\
&\hspace{20mm}+D_{21}^{z}(S_{s}^{x}(\bm{R},A)S_{s}^{y}(\bm{R}+\bm{a}_{21},A)-S_{s}^{y}(\bm{R},A)S_{s}^{x}(\bm{R}+\bm{a}_{21},A)) \nonumber \\
&\hspace{20mm}+D_{31}^{z}(S_{s}^{x}(\bm{R},A)S_{s}^{y}(\bm{R}+\bm{a}_{31},A)-S_{s}^{y}(\bm{R},A)S_{s}^{x}(\bm{R}+\bm{a}_{31},A)) \nonumber \\
&\hspace{20mm}+D_{32}^{z}(S_{s}^{x}(\bm{R},A)S_{s}^{y}(\bm{R}+\bm{a}_{32},A)-S_{s}^{y}(\bm{R},A)S_{s}^{x}(\bm{R}+\bm{a}_{32},A)) \nonumber \\
&\hspace{20mm}-(A\leftrightarrow B) , \\
&H_{J'}=J'\sum_{i}\bm{S}_{i,{\rm u}}\cdot\bm{S}_{i,{\rm d}}, \\
&H_{J} 
=-\sum_{\bm{R},s={\rm u,d}}J_{0}\bm{S}_{s}(\bm{R},A)\cdot\bm{S}_{s}(\bm{R},B)+J_{1}\bm{S}_{s}(\bm{R},A)\cdot\bm{S}_{s}(\bm{R}+\bm{a}_{1},B) \nonumber \\
&\hspace{20mm}+J_{2}\bm{S}_{s}(\bm{R},A)\cdot\bm{S}_{s}(\bm{R}+\bm{a}_{2},B)+J_{3}\bm{S}_{s}(\bm{R},A)\cdot\bm{S}_{s}(\bm{R}+\bm{a}_{3},B) ,\label{eq:bond_d} \\
&H_{\rm XY} 
=J_{-}\sum_{\bm{R}}
\bar{D}_{1}^{y}(S_{\rm u}^{x}(\bm{R},A)S_{\rm d}^{x}(\bm{R}+\bm{a}_{1},A)-S_{\rm u}^{y}(\bm{R},A)S_{\rm d}^{y}(\bm{R}+\bm{a}_{1},A)) \nonumber \\
&\hspace{18.5mm}+\bar{D}_{2}^{y}(S_{\rm u}^{x}(\bm{R},A)S_{\rm d}^{x}(\bm{R}+\bm{a}_{2},A)-S_{\rm u}^{y}(\bm{R},A)S_{\rm d}^{y}(\bm{R}+\bm{a}_{2},A)) \nonumber \\
&\hspace{18.5mm}+\bar{D}_{3}^{y}(S_{\rm u}^{x}(\bm{R},A)S_{\rm d}^{x}(\bm{R}+\bm{a}_{3},A)-S_{\rm u}^{y}(\bm{R},A)S_{\rm d}^{y}(\bm{R}+\bm{a}_{3},A)) \nonumber \\
&\hspace{18.5mm}+\bar{D}_{21}^{y}(S_{\rm u}^{x}(\bm{R},A)S_{\rm d}^{x}(\bm{R}+\bm{a}_{21},A)-S_{\rm u}^{y}(\bm{R},A)S_{\rm d}^{y}(\bm{R}+\bm{a}_{21},A)) \nonumber \\
&\hspace{18.5mm}+\bar{D}_{31}^{y}(S_{\rm u}^{x}(\bm{R},A)S_{\rm d}^{x}(\bm{R}+\bm{a}_{31},A)-S_{\rm u}^{y}(\bm{R},A)S_{\rm d}^{y}(\bm{R}+\bm{a}_{31},A)) \nonumber \\
&\hspace{18.5mm}+\bar{D}_{32}^{y}(S_{\rm u}^{x}(\bm{R},A)S_{\rm d}^{x}(\bm{R}+\bm{a}_{32},A)-S_{\rm u}^{y}(\bm{R},A)S_{\rm d}^{y}(\bm{R}+\bm{a}_{32},A)) \nonumber \\
&\hspace{18.5mm}-({\rm u}\leftrightarrow{\rm d}) \nonumber \\
&\hspace{18.5mm}-(A\leftrightarrow B) , \\
&H_{\rm \Gamma} 
=\Gamma\sum_{\bm{R}}
\bar{D}_{1}^{x}(S_{\rm u}^{x}(\bm{R},A)S_{\rm d}^{y}(\bm{R}+\bm{a}_{1},A)+S_{\rm u}^{y}(\bm{R},A)S_{\rm d}^{x}(\bm{R}+\bm{a}_{1},A)) \nonumber \\
&\hspace{14mm}+\bar{D}_{2}^{x}(S_{\rm u}^{x}(\bm{R},A)S_{\rm d}^{y}(\bm{R}+\bm{a}_{2},A)+S_{\rm u}^{y}(\bm{R},A)S_{\rm d}^{x}(\bm{R}+\bm{a}_{2},A)) \nonumber \\
&\hspace{14mm}+\bar{D}_{3}^{x}(S_{\rm u}^{x}(\bm{R},A)S_{\rm d}^{y}(\bm{R}+\bm{a}_{3},A)+S_{\rm u}^{y}(\bm{R},A)S_{\rm d}^{x}(\bm{R}+\bm{a}_{3},A)) \nonumber \\
&\hspace{14mm}+\bar{D}_{21}^{x}(S_{\rm u}^{x}(\bm{R},A)S_{\rm d}^{y}(\bm{R}+\bm{a}_{21},A)+S_{\rm u}^{y}(\bm{R},A)S_{\rm d}^{x}(\bm{R}+\bm{a}_{21},A)) \nonumber \\
&\hspace{14mm}+\bar{D}_{31}^{x}(S_{\rm u}^{x}(\bm{R},A)S_{\rm d}^{y}(\bm{R}+\bm{a}_{31},A)+S_{\rm u}^{y}(\bm{R},A)S_{\rm d}^{x}(\bm{R}+\bm{a}_{31},A)) \nonumber \\
&\hspace{14mm}+\bar{D}_{32}^{x}(S_{\rm u}^{x}(\bm{R},A)S_{\rm d}^{y}(\bm{R}+\bm{a}_{32},A)+S_{\rm u}^{y}(\bm{R},A)S_{\rm d}^{x}(\bm{R}+\bm{a}_{32},A)) \nonumber \\
&\hspace{14mm}-({\rm u}\leftrightarrow{\rm d}) \nonumber \\
&\hspace{14mm}-(A\leftrightarrow B) , \\
&H_{\kappa} =-\kappa \sum_{i,s={\rm u,d}}(S_{i,s}^{z})^{2},
\end{align}
\end{widetext}
where $\bm{S}_{i,{\rm u}}$ and $\bm{S}_{i,{\rm d}}$ are the operators of up and down spins localized at the site $i$, respectively.
The operator $\bm{S}_{s}(\bm{R},X)$ $(s={\rm u,d})$ is the another expression of $\bm{S}_{i,s}$ $(s={\rm u,d})$ for the site $i$ on $X$ sublattice in the unit cell labeled by the lattice vector $\bm{R}$.
The vector $\bm{a}_{ij}$ is defined as $\bm{a}_{ij}=\bm{a}_{i}-\bm{a}_{j}$.
The DM vectors $\bm{D}_{i}$ ($\bm{D}_{ij}$) is written as $\bm{D}_{i}=D(\bm{d}^{1}_{i}(\bm{R})\times\bm{d}^{2}_{i}(\bm{R}))/|\bm{d}^{1}_{i}(\bm{R})\times\bm{d}^{2}_{i}(\bm{R})|$ $\left( \bm{D}_{ij}=D(\bm{d}^{1}_{ij}(\bm{R})\times\bm{d}^{2}_{ij}(\bm{R}))/|\bm{d}^{1}_{ij}(\bm{R})\times\bm{d}^{2}_{ij}(\bm{R})|\right)$, where $\bm{d}^{1,2}_{i}(\bm{R})$ $(\bm{d}^{1,2}_{ij}(\bm{R}))$ are the two nearest neighbor bond vectors traversed between sites $(\bm{R},A)$ and $(\bm{R}+\bm{a}_{i},A)$ ($(\bm{R},A)$ and $(\bm{R}+\bm{a}_{ij},A)$).
Figure \ref{fig:DMvector} shows the details of $\bm{d}^{1,2}_{i}(\bm{R})$.
The bond dependence of the Heisenberg interaction (\ref{eq:bond_d}) originates from 
pressure-induced lattice distortion.
Here $\bar{\bm{D}}_{i}$ and $\bar{\bm{D}}_{ij}$ are written as $\bar{\bm{D}}_{i}=\bm{D}_{i}/D$ and $\bar{\bm{D}}_{ij}=\bm{D}_{ij}/D$, respectively.

\begin{center}
\bf{Appendix B: Simplified expressions for the topological invariants of the diamond lattice system}
\end{center}

If a system has inversion symmetry, the expression of the topological invariants for topological insulators can be greatly simplified~\cite{Fu07b}.
The same discussion can be applied to magnon systems.
In this part, by using inversion symmetry, we analytically calculate the topological invariants for diamond lattice system.
The Hamiltonian (\ref{eq:H_FKM}) satisfies the following inversion symmetry:
\begin{align}
R\Sigma_{z}H(\bm{k})-\Sigma_{z}H(-\bm{k})R=0,
\end{align}
where $R$ is an inversion operator defined as $R:=1_{2}\otimes 1_{2}\otimes \sigma_{x}$.
Following the discussion in Ref.~\cite{Fu07b}, topological invariants for 3D topological magnon systems with inversion symmetry can be written as
\begin{align}
&(-1)^{\nu_{0}}=\prod_{n_{1}=0,1;n_{2}=0,1;n_{3}=0,1}\delta_{m=(n_{1}n_{2}n_{3})}, \nonumber \\
&(-1)^{\nu_{i}}=\prod_{n_{i}=1;n_{j\neq i}=0,1}\delta_{m=(n_{1}n_{2}n_{3})},
\end{align}
where $i=x,y$, and $z$.
Since $\Sigma_{z}H(\bm{k})$ commutes with the inversion operator $R$ at TRIM: $\bm{\Gamma}_{m}=\pi(n_{1},n_{2},n_{3})$, an energy eigenvector $\bm{\Psi}_{n,1,+}(\bm{\Gamma}_{m})$ can be chosen to be an eigenvector of $R$. Here, we write the eigenvalue of $R$ as $\xi_{n}(\bm{\Gamma}_{m})$. 
Then, $\delta_{m=(n_{1}n_{2}n_{3})}$ is defined as the product of $\xi_{n}(\bm{\Gamma}_{m})$ over the bands below the fictitious Fermi energy:
\begin{align}
\delta_{m=(n_{1}n_{2}n_{3})}=\prod_{n,(E_{n}({\bm k}) \leq \epsilon_{0})}\xi_{n}(\bm{\Gamma}_{m})\label{eq:delta}
\end{align}
Now, we assume that the fictitious Fermi energy lies between two bands in the particle space.
In this case, $\delta_{m=(n_{1}n_{2}n_{3})}$ is an eigenvalue of  $R$ with eigenvector of the lower band $\bm{\Psi}_{2,1,+}(\bm{\Gamma}_{m})$.
In the following, we obtain the expression of $\bm{\Psi}_{2,1,+}(\bm{\Gamma}_{m})$.
At the TRIM, the Hamiltonian of the diamond lattice system (\ref{eq:Ham_gen}) is written as
\begin{align}
H(\bm{k})=\left(
\begin{array}{cccc}
h_{1}(\bm{\Gamma})  &0 &0 &J'1_{2}  \\
0 &h_{1}^{*}(\bm{\Gamma}) &J'1_{2} &0   \\
0  &-J'1_{2} &-h_{1}^{*}(\bm{\Gamma}) &0  \\
-J'1_{2} &0 &0&-h_{1}(\bm{\Gamma}) 
\end{array} 
\right).
\end{align}
The eigenvectors of the matrix $h_{1}(\bm{\Gamma})$ are written as:
\begin{align}
\bm{\phi}_{\pm}(\bm{\Gamma})=\frac{1}{\sqrt{2}}
\left(
\begin{array}{cc}
1   \\
\mp {\rm sgn}[\gamma(\bm{\Gamma})]     \\
\end{array} 
\right),
\end{align}
where the corresponding eigenvalues are $\epsilon_{\pm}(\bm{\Gamma})=d_{0}\pm|d_{1}(\bm{\Gamma})|$. 
Here, $\gamma(\bm{k})$ is defined as $\gamma(\bm{k})= -d_{1}(\bm{k})-{\rm i} d_{2}(\bm{k})$.
The concrete expressions of $d_{0}$ and $d_{i}(\bm{k})$ $(i=1,\cdots, 5)$ are given in the main text.
We note that imaginary part of $\gamma(\bm{k})$, i.e. $d_{2}(\bm{k})$, vanishes at TRIM.
If we write the eigenvector of $\Sigma_{z}H(\bm{\Gamma})$ as
\begin{align}
\bm{\Phi}_{\pm}(\bm{\Gamma})=
\left(
\begin{array}{cccc}
\alpha\bm{\phi}_{\pm}(\bm{\Gamma})   \\
0    \\
0    \\
\beta\bm{\phi}_{\pm}(\bm{\Gamma}) 
\end{array} 
\right),
\end{align}
the eigenvalue equation can be written as follows:

\begin{widetext}
\begin{align}
\left(
\begin{array}{cccc}
h_{1}(\bm{\Gamma})  &0 &0 &J'1_{2}  \\
0 &h_{1}^{*}(\bm{\Gamma}) &J'1_{2} &0   \\
0  &-J'1_{2} &-h_{1}^{*}(\bm{\Gamma}) &0  \\
-J'1_{2} &0 &0&-h_{1}(\bm{\Gamma}) 
\end{array} 
\right)
\left(
\begin{array}{cccc}
\alpha\bm{\phi}_{\pm}(\bm{\Gamma})   \\
0    \\
0    \\
\beta\bm{\phi}_{\pm}(\bm{\Gamma}) 
\end{array} 
\right)
=E_{\pm}(\bm{\Gamma})
\left(
\begin{array}{cccc}
\alpha\bm{\phi}_{\pm}(\bm{\Gamma})   \\
0    \\
0    \\
\beta\bm{\phi}_{\pm}(\bm{\Gamma}) 
\end{array} 
\right).
\end{align}
\end{widetext}
This results in the following relation:
\begin{align}
\left(
\begin{array}{cc}
\epsilon_{\pm}(\bm{\Gamma})-E_{\pm}(\bm{\Gamma})  &J' \\
-J' & -\epsilon_{\pm}(\bm{\Gamma})-E_{\pm}(\bm{\Gamma})   \\
\end{array} 
\right)
\left(
\begin{array}{cc}
\alpha\\
\beta  \\
\end{array} 
\right)
=\left(
\begin{array}{cc}
0\\
0\\
\end{array} 
\right).
\end{align}
In order for the equation to have a nontrivial solution, the determinant of the matrix in the left hand side must be zero.
Then, we obtain the following equation:
\begin{align}
E_{\pm}^{2}(\bm{\Gamma})-\epsilon_{\pm}^{2}(\bm{\Gamma})+J'^{2}=0.
\end{align}
Therefore, we obtain the energy eigenvalues in the particle space:
\begin{align}
E_{\pm}(\bm{\Gamma})=\sqrt{\epsilon_{\pm}^{2}(\bm{\Gamma})-J'^{2}}.\label{eq:E_TRIM}
\end{align}
We can see that $E_{+}(\bm{\Gamma})$ and $E_{-}(\bm{\Gamma})$ correspond to the energies of the top and bottom bands, respectively.
Now we focus on the bottom band.
By using $\theta(\bm{\Gamma})$ defined as
\begin{align}
\tanh{(\theta(\bm{\Gamma}))}:=\frac{E_{-}(\bm{\Gamma})-\epsilon_{-}(\bm{\Gamma})}{J'},
\end{align}
coefficients $\alpha$ and $\beta$ of the normalized wave function of the bottom band are written as
$\alpha=\cosh{(\theta(\bm{\Gamma}))}$ and $\beta=\sinh{(\theta(\bm{\Gamma}))}$, respectively.
Then, the eigenvector of the bottom band $\bm{\Psi}_{2,1,+}(\bm{\Gamma})$ are given by:
\begin{align}
\bm{\Psi}_{2,1,+}(\bm{\Gamma})=
\left(
\begin{array}{cccc}
\cosh{(\theta(\bm{\Gamma}))}\bm{\phi}_{-}(\bm{\Gamma})   \\
0    \\
0    \\
\sinh{(\theta(\bm{\Gamma}))}\bm{\phi}_{-}(\bm{\Gamma}) 
\end{array} 
\right).
\end{align}
We can see that this is an eigenvector of $R$ with eigenvalue ${\rm sgn}[\gamma(\bm{\Gamma})]$.
Then, $\delta_{m=(n_{1}n_{2}n_{3})}$ in Eq.~(\ref{eq:delta}) can be written as
\begin{align}
\delta_{m=(n_{1}n_{2}n_{3})}&={\rm sgn}[\gamma(\bm{\Gamma}_{m})] \nonumber \\
&={\rm sgn}[J_{0}+J_{1}e^{{\rm i}\Gamma_{m}^{x}}+J_{2}e^{{\rm i}\Gamma_{m}^{y}}+J_{3}e^{{\rm i}\Gamma_{m}^{z}}].
\end{align}
By using this, the strong index for the diamond lattice system is given by
\begin{widetext}
\begin{align}
(-1)^{\nu_{0}}
&={\rm sgn}[(J_{0}-J_{1}+J_{2}+J_{3})(J_{0}-J_{1}-J_{2}+J_{3})(J_{0}-J_{1}+J_{2}-J_{3})(J_{0}-J_{1}-J_{2}-J_{3})] \nonumber \\
&\times{\rm sgn}[(J_{0}+J_{1}+J_{2}+J_{3})(J_{0}+J_{1}-J_{2}+J_{3})(J_{0}+J_{1}+J_{2}-J_{3})(J_{0}+J_{1}-J_{2}-J_{3})].
\end{align}
The other three indices are written as follows:
\begin{align}
&(-1)^{\nu_{x}}
={\rm sgn}[(J_{0}-J_{1}+J_{2}+J_{3})(J_{0}-J_{1}-J_{2}+J_{3})(J_{0}-J_{1}+J_{2}-J_{3})(J_{0}-J_{1}-J_{2}-J_{3})],
\\
&(-1)^{\nu_{y}}
={\rm sgn}[(J_{0}+J_{1}-J_{2}+J_{3})(J_{0}-J_{1}-J_{2}+J_{3})(J_{0}+J_{1}-J_{2}-J_{3})(J_{0}-J_{1}-J_{2}-J_{3})] ,
\\
&(-1)^{\nu_{z}}
={\rm sgn}[(J_{0}+J_{1}+J_{2}-J_{3})(J_{0}-J_{1}+J_{2}-J_{3})(J_{0}+J_{1}-J_{2}-J_{3})(J_{0}-J_{1}-J_{2}-J_{3})] .
\end{align}
\end{widetext}

\begin{center}
\bf{Appendix C: The analytical expressions of energy eigenvalues}
\end{center}
In the case of $\Gamma=0$, the energy spectrum of the Hamiltonian of the diamond lattice system (\ref{eq:Ham_di}) can be obtained analytically.
By using the unitary matrix $U(\bm{k})$ (details of $U(\bm{k})$ are shown later), $\Sigma_{z}H(\bm{k})$ can be written as follows:
\begin{align}
U^{\dagger}(\bm{k})\Sigma_{z}H(\bm{k}) U(\bm{k})=
\left(
\begin{array}{cc}
0 &Q_{1}(\bm{k})   \\
Q_{2}(\bm{k})  &0   \\
\end{array} 
\right), \label{eq:Ham_OD}
\end{align}
where $Q_{n}(\bm{k})$ $(n=1,2)$ are defined by
\begin{align}
Q_{n}(\bm{k})=
\left(
\begin{array}{cccc}
-d_{0} &0 &\lambda_{n,+}(\bm{k}) &0  \\
0 &-d_{0} &0 &\lambda_{n,-}(\bm{k})  \\
\lambda_{n,+}^{*}(\bm{k}) &0 &-d_{0} &0  \\
0 &\lambda_{n,-}^{*}(\bm{k}) &0 &-d_{0}  \\
\end{array} 
\right).
\end{align}
Here $\lambda_{n,\pm}(\bm{k})$ is written as
\begin{align}
\lambda_{n,\pm}(\bm{k})=-{\rm i} d_{4}(\bm{k})+(-1)^{n}J'\pm\sqrt{d_{5}^{2}(\bm{k})+|\gamma(\bm{k})|^{2}} .
\end{align}
We note that the matrix $U^{\dagger}(\bm{k})(\Sigma_{z}H(\bm{k}))^{2} U(\bm{k})$, whose eigenvalues are equal to the square of the ones of $\Sigma_{z}H(\bm{k})$, is the block diagonal:
\begin{align}
U^{\dagger}(\bm{k})(\Sigma_{z}H(\bm{k}))^{2} U(\bm{k})
&=(U^{\dagger}(\bm{k})\Sigma_{z}H(\bm{k}) U(\bm{k}))^{2} \nonumber \\
&=\left(
\begin{array}{cc}
Q_{1}(\bm{k})Q_{2}(\bm{k}) &0   \\
0  &Q_{2}(\bm{k})Q_{1}(\bm{k})   \\
\end{array} 
\right).
\end{align}
This suggests that the square of the energy eigenvalue $E^{2}(\bm{k})$ can be obtained as the eigenvalues of $Q_{1}(\bm{k})Q_{2}(\bm{k})$:
\begin{widetext}
\begin{align}
Q_{1}(\bm{k})Q_{2}(\bm{k})=
\left(
\begin{array}{cccc}
d_{0}^{2}+\lambda_{1,+}(\bm{k})\lambda_{2,+}^{*}(\bm{k}) &0 &-d_{0}(\lambda_{1,+}(\bm{k})+\lambda_{2,+}(\bm{k})) &0  \\
0 &d_{0}^{2}+\lambda_{1,-}(\bm{k})\lambda_{2,-}^{*}(\bm{k}) &0 &-d_{0}(\lambda_{1,-}(\bm{k})+\lambda_{2,-}(\bm{k}))  \\
-d_{0}(\lambda_{1,+}^{*}(\bm{k})+\lambda_{2,+}^{*}(\bm{k})) &0 &d_{0}^{2}+\lambda_{1,+}^{*}(\bm{k})\lambda_{2,+}(\bm{k}) &0  \\
0 &-d_{0}(\lambda_{1,-}^{*}(\bm{k})+\lambda_{2,-}^{*}(\bm{k})) &0 &d_{0}^{2}+\lambda_{1,-}^{*}(\bm{k})\lambda_{2,-}(\bm{k})  \\
\end{array} 
\right).
\end{align}
Since this matrix  can be divided into two parts, the eigenvalues are calculated from the following equation:
\begin{align}
\left|
\begin{array}{cc}
d_{0}^{2}+\lambda_{1,\pm}(\bm{k})\lambda_{2,\pm}^{*}(\bm{k})-E^{2}(\bm{k}) &-d_{0}(\lambda_{1,\pm}(\bm{k})+\lambda_{2,\pm}(\bm{k})) \\
-d_{0}(\lambda_{1,\pm}^{*}(\bm{k})+\lambda_{2,\pm}^{*}(\bm{k})) &d_{0}^{2}+\lambda_{1,\pm}^{*}(\bm{k})\lambda_{2,\pm}(\bm{k})-E^{2}(\bm{k}) \\
\end{array} 
\right|=0.
\end{align}
Then, one has
\begin{align}
E^{2}(\bm{k}) =d_{0}^{2} +{\rm Re}[\lambda_{1,\rho}(\bm{k})\lambda_{2,\rho}^{*}(\bm{k}) ] +\rho' \sqrt{(d_{0}^{2}+ {\rm Re}[\lambda_{1,\rho}(\bm{k})\lambda_{2,\rho}^{*}(\bm{k}) ])^{2}+d_{0}^{2}|\lambda_{1,\rho}(\bm{k})+\lambda_{2,\rho}(\bm{k})|^{2}}
\hspace{5mm}(\rho,\rho'=\pm).
\end{align}
Finally, the eigenvalues of $\Sigma_{z}H(\bm{k})$ are written as follows:
\begin{align}
E_{\rho_{1},\rho_{2},\rho_{3}}(\bm{k}) =
\rho_{1}\sqrt{
d_{0}^{2} +{\rm Re}[\lambda_{1,\rho_{2}}(\bm{k})\lambda_{2,\rho_{2}}^{*}(\bm{k}) ]+ \rho_{3} \sqrt{(d_{0}^{2}+ {\rm Re}[\lambda_{1,\rho_{2}}(\bm{k})\lambda_{2,\rho_{2}}^{*}(\bm{k}) ])^{2}+d_{0}^{2}|\lambda_{1,\rho_{2}}(\bm{k})+\lambda_{2,\rho_{2}}(\bm{k})|^{2}}
},
\end{align}
\end{widetext}
where $\rho_{1},\rho_{2},\rho_{3}=\pm$.
We note that due to the pseudo-time-reversal and inversion symmetry, these eigenvalues are doubly degenerate and  satisfy $E_{\rho_{1},+,\rho_{3}}(\bm{k})=E_{\rho_{1},-,\rho_{3}}(\bm{k})$.

In the following, we show how to construct the matrix $U(\bm{k})$.
This is written as the product of three unitary matrices $U_{1},U_{2},$ and $U_{3}(\bm{k})$, i.e.  $U(\bm{k})=U_{1}U_{2}U_{3}(\bm{k})$.
The first matrix $U_{1}$ is the one which diagonalizes $C:=\sigma_{y}\otimes \sigma_{y}\otimes 1_{2}$.
Since  $C$ anticommutes with $\Sigma_{z}H(\bm{k})$,  $U_{1}$ makes $\Sigma_{z}H(\bm{k})$ off diagonal:
\begin{align}
U_{1}^{\dagger}\Sigma_{z}H(\bm{k}) U_{1}=
\left(
\begin{array}{cc}
0 &R_{1}(\bm{k})   \\
R_{2}(\bm{k})  &0   \\
\end{array} 
\right),
\end{align}
where
\begin{widetext}
\begin{align}
R_{1}(\bm{k})=
\left(
\begin{array}{cccccccc}
-d_{0}+d_{5}(\bm{k})+J' &\gamma(\bm{k}) &{\rm i} d_{4}(\bm{k}) &0   \\
-\gamma^{*}(\bm{k}) &-d_{0}-d_{5}(\bm{k})+J' &0 &-{\rm i} d_{4}(\bm{k})   \\
-{\rm i} d_{4}(\bm{k})&0 &-d_{0}-d_{5}(\bm{k})-J' &\gamma(\bm{k})   \\
0 &{\rm i} d_{4}(\bm{k}) &-\gamma^{*}(\bm{k}) &-d_{0}+d_{5}(\bm{k})-J'    \\
\end{array} 
\right),
\end{align}
\begin{align}
R_{2}(\bm{k})=
\left(
\begin{array}{cccccccc}
-d_{0}+d_{5}(\bm{k})-J' &\gamma(\bm{k}) &{\rm i} d_{4}(\bm{k}) &0   \\
-\gamma^{*}(\bm{k}) &-d_{0}-d_{5}(\bm{k})-J' &0 &-{\rm i} d_{4}(\bm{k})   \\
-{\rm i} d_{4}(\bm{k}) &0 &-d_{0}-d_{5}(\bm{k})+J' &\gamma(\bm{k})   \\
0 &{\rm i} d_{4}(\bm{k}) &-\gamma^{*}(\bm{k}) &-d_{0}+d_{5}(\bm{k})+J'    \\
\end{array} 
\right).
\end{align}
\end{widetext}
The second matrix is given by $U_{2}=1_{2}\otimes u$. Here $u$ is a $4\times 4$ matrix and diagonalizes $c:=\sigma_{x}\otimes \sigma_{z}$ which anticommutes with $d_{0} 1_{4}+R_{1,2}(\bm{k})$.
By the unitary transformation using $u$, $d_{0} 1_{4}+R_{1,2}(\bm{k})$ becomes off diagonal:
\begin{align}
u^{\dagger}(d_{0} 1_{4}+R_{1}(\bm{k}))u=
\left(
\begin{array}{cc}
0 &r_{11}(\bm{k})    \\
r_{12}(\bm{k})  &0    \\
\end{array} 
\right),
\end{align}
\begin{align}
u^{\dagger}(d_{0} 1_{4}+R_{2}(\bm{k}))u=
\left(
\begin{array}{cc}
0 &r_{21}(\bm{k})    \\
r_{22}(\bm{k})  &0    \\
\end{array} 
\right),
\end{align}
where
\begin{align}
&r_{11}(\bm{k}) =r_{12}^{\dagger}(\bm{k}) \nonumber \\
&=
\left(
\begin{array}{cc}
d_{5}(\bm{k})-{\rm i} d_{4}(\bm{k})-J'  & \gamma^{*}(\bm{k}) \\
\gamma(\bm{k})  &-d_{5}(\bm{k})-{\rm i} d_{4}(\bm{k})-J'    \\
\end{array} 
\right),
\end{align}
\begin{align}
&r_{21}(\bm{k}) =r_{22}^{\dagger}(\bm{k}) \nonumber \\
&=
\left(
\begin{array}{cc}
d_{5}(\bm{k})-{\rm i} d_{4}(\bm{k})+J'  & \gamma^{*}(\bm{k}) \\
\gamma(\bm{k})  &-d_{5}(\bm{k})-{\rm i} d_{4}(\bm{k})+J'    \\
\end{array} 
\right).
\end{align}
Since $r_{ij}(\bm{k})$ $(i,j=1,2)$ commute with each other, they are diagonalized by the same unitary matrix $v(\bm{k})$ as follows:
\begin{align}
&v^{\dagger}(\bm{k})r_{n1}(\bm{k}) v(\bm{k})=
\left(
\begin{array}{cc}
\lambda_{n,+}(\bm{k})  & 0 \\
0  &\lambda_{n,-}(\bm{k})    \\
\end{array} 
\right), \\
&v^{\dagger}(\bm{k})r_{n2}(\bm{k}) v(\bm{k})=
\left(
\begin{array}{cc}
\lambda_{n,+}^{*}(\bm{k})  & 0 \\
0  &\lambda_{n,-}^{*}(\bm{k})    \\
\end{array} 
\right)\hspace{5mm} (n=1,2).
\end{align}
By using $v(\bm{k})$, the third matrix $U_{3}(\bm{k})$ is defined as $U_{3}(\bm{k})=1_{4}\otimes v(\bm{k})$.
Now we can construct the matrix  $U(\bm{k})$ and obtain the unitary-transformed Hamiltonian Eq.~(\ref{eq:Ham_OD}).

\begin{center}
\bf{Appendix D: Second model of 3D topological magnon systems: perovskite lattice system}
\end{center}

\begin{figure}[H]
\centering
  \includegraphics[width=7cm]{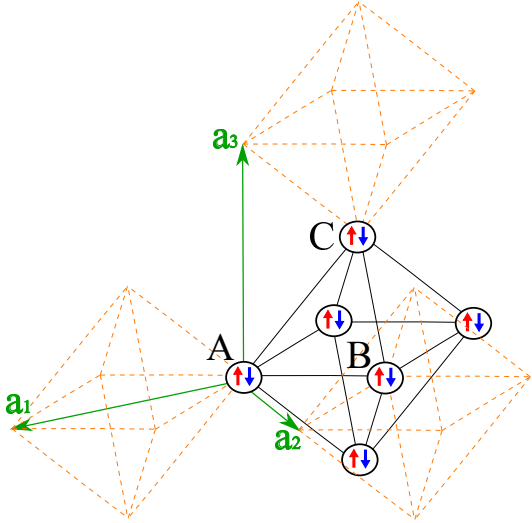}
\caption{Perovskite lattice system having two spins at each lattice site. Three sublattices are indicated by $A,B$, and $C$. The vectors $\bm{a}_{i}$ $(i=1,2,3)$ are the lattice prime vectors. We take $\bm{a}_{1}=(1,0,0),\bm{a}_{2}=(0,1,0)$, and $\bm{a}_{3}=(0,0,1)$.
}\label{fig:cell_pe}
\end{figure}

The second example of 3D magnon topological systems is a system in which two spins with opposite directions are localized at each site of perovskite lattice.
The system is depicted in Fig.~\ref{fig:cell_pe}.
The Hamiltonian of the system are written as
\begin{align}
H=H_{\rm DM}+H_{J}+H_{\rm XY}+H_{J'}+H_{\kappa}.  \label{eq:Ham_pe}
\end{align}
Here, $H_{\rm DM},H_{J}$ are the DM and ferromagnetic Heisenberg interactions between nearest-neighbor spins aligned in the same direction. The third term $H_{\rm XY}$ is the anisotropic XY interaction between nearest-neighbor spins pointing in opposite directions.
The other terms $H_{J'}$ and $H_{\kappa}$ are the antiferromagnetic interaction between spins at the same lattice site and the single ion anisotropy, respectively. We note that the single-ion anisotropy is introduced to open a band gap.
\noindent
The explicit forms of the terms are given by
\begin{align}
&H_{\rm DM}=\sum_{\langle i,j\rangle,s={\rm u,d}} \bm{D}_{ij}\cdot (\bm{S}_{i,s}\times\bm{S}_{j,s}), \\
&H_{J}=-J\sum_{\langle i,j\rangle,s={\rm u,d}}\bm{S}_{i,s}\cdot\bm{S}_{j,s},  \\
&H_{J'}=J'\sum_{i}\bm{S}_{i,{\rm u}}\cdot\bm{S}_{i,{\rm d}},  \\
&H_{\kappa}=-\sum_{X=A,B,C}\sum_{i\in X,s}\kappa_{X}(S_{i,s}^{z})^{2}.
\end{align}
Here, we take the directions of ordered spins as $\pm\bm{s}=\pm(s_{1},s_{2},s_{3})$ and write corresponding spin operators at site $i$ as $\bm{S}_{i,{\rm u}},\bm{S}_{i,{\rm d}}$, respectively. The Hamiltonian of anisotropic XY interaction is written as follows:
\begin{widetext}
\begin{align}
H_{\rm XY} 
&=\sum_{\bm{R}}\bm{S}_{\rm u}(\bm{R},A)J_{\rm XY}\left[\bm{S}_{\rm d}(\bm{R},B)+\bm{S}_{\rm d}(\bm{R}+\bm{a}_{1},B)+\bm{S}_{\rm d}(\bm{R}-\bm{a}_{2},B)+\bm{S}_{\rm d}(\bm{R}+\bm{a}_{12},B)\right] \nonumber \\
&\hspace{6mm}+\bm{S}_{\rm u}(\bm{R},A)J_{\rm XY}\left[\bm{S}_{\rm d}(\bm{R},C)+\bm{S}_{\rm d}(\bm{R}+\bm{a}_{1},C)+\bm{S}_{\rm d}(\bm{R}-\bm{a}_{3},C)+\bm{S}_{\rm d}(\bm{R}+\bm{a}_{13},C)\right] \nonumber \\
&\hspace{6mm}+\bm{S}_{\rm u}(\bm{R},B)J_{\rm XY}\left[\bm{S}_{\rm d}(\bm{R},C)+\bm{S}_{\rm d}(\bm{R}+\bm{a}_{2},C)+\bm{S}_{\rm d}(\bm{R}-\bm{a}_{3},C)+\bm{S}_{\rm d}(\bm{R}+\bm{a}_{23},C)\right] \nonumber \\
&\hspace{6mm}+\bm{S}_{\rm d}(\bm{R},A)J_{\rm YX}\left[\bm{S}_{\rm u}(\bm{R},B)+\bm{S}_{\rm u}(\bm{R}+\bm{a}_{1},B)+\bm{S}_{\rm u}(\bm{R}-\bm{a}_{2},B)+\bm{S}_{\rm u}(\bm{R}+\bm{a}_{12},B)\right] \nonumber \\
&\hspace{6mm}+\bm{S}_{\rm d}(\bm{R},A)J_{\rm YX}\left[\bm{S}_{\rm u}(\bm{R},C)+\bm{S}_{\rm u}(\bm{R}+\bm{a}_{1},C)+\bm{S}_{\rm u}(\bm{R}-\bm{a}_{3},C)+\bm{S}_{\rm u}(\bm{R}+\bm{a}_{13},C)\right] \nonumber \\
&\hspace{6mm}+\bm{S}_{\rm d}(\bm{R},B)J_{\rm YX}\left[\bm{S}_{\rm u}(\bm{R},C)+\bm{S}_{\rm u}(\bm{R}+\bm{a}_{2},C)+\bm{S}_{\rm u}(\bm{R}-\bm{a}_{3},C)+\bm{S}_{\rm u}(\bm{R}+\bm{a}_{23},C)\right] . \nonumber
\end{align}
\end{widetext}
The diagonal matrices $J_{\rm XY}$ and $J_{\rm YX}$ are $3\times 3$ matrices and written as $J_{\rm XY}={\rm diag}[J_{x},J_{y},0]$ and $J_{\rm YX}={\rm diag}[J_{y},J_{x},0]$, respectively. 
Assuming the ligands are located on the center of the octahedra of the perovskite lattice,
we difine the DM vector $\bm{D}_{ij}$ for the nearest neighbor sites $i,j$ as $\bm{D}_{ij}=D(\bm{d}_{i}\times\bm{d}_{j})/|\bm{d}_{i}\times\bm{d}_{j}|$, where $\bm{d}_{i}$ and $\bm{d}_{j}$ are the vectors from sites $i,j$ to the nearest ligand.

Here, the most general Hamiltonian satisfying Eq.~(\ref{eq:Commutation}) takes the form: 
\begin{align}
H(\bm{k})=
\left(
\begin{array}{cccc}
h_{1}(\bm{k}) &h_{2}(\bm{k}) &\Delta_{2}(\bm{k}) &\Delta_{1}(\bm{k}) \\
h_{2}^{\dagger}(\bm{k}) &h_{1}^{*}(-\bm{k}) &\Delta_{1}^{*}(-\bm{k})&-\Delta_{2}^{\dagger}(\bm{k}) \\
\Delta_{2}^{\dagger}(\bm{k}) &\Delta_{1}^{*}(-\bm{k}) &h_{1}^{*}(-\bm{k}) &h_{2}^{*}(-\bm{k}) \\
\Delta_{1}(\bm{k}) &-\Delta_{2}(\bm{k}) &h_{2}^{T}(-\bm{k}) &h_{1}(\bm{k}) \\
\end{array} 
\right)
\label{eq:Ham_gen}
\end{align}
where $h_i(\bm{k})$ and $\Delta_i(\bm{k})$ for $i=1, 2$ are $3 \times 3$ matrices and satisfy $h_{1}^{\dagger}(\bm{k})=h_{1}(\bm{k}),\Delta_{1}^{\dagger}(\bm{k})=\Delta_{1}(\bm{k}),h_{2}^{T}(\bm{k})=-h_{2}(-\bm{k}),$ and $\Delta_{2}^{T}(\bm{k})=\Delta_{2}(-\bm{k})$.
Applying the Holstein-Primakoff transformation, the Hamiltonian (\ref{eq:Ham_pe}) boils down to the same form as Eq.~(\ref{eq:Ham_gen}). 
We take the operators $\bm{b}_{\uparrow}^{\dagger}(\bm{k})$ and $\bm{b}_{\downarrow}^{\dagger}(\bm{k})$ in Eq.~(\ref{eq:opvec}) as
\begin{align}
&\bm{b}_{\uparrow}^{\dagger}(\bm{k})=[b_{\rm u}^{\dagger}(\bm{k},A),b_{\rm u}^{\dagger}(\bm{k},B),b_{\rm u}^{\dagger}(\bm{k},C)] , \\
&\bm{b}_{\downarrow}^{\dagger}(\bm{k})=[b_{\rm d}^{\dagger}(\bm{k},A),b_{\rm d}^{\dagger}(\bm{k},B),b_{\rm d}^{\dagger}(\bm{k},C)] .
\end{align}
Writing the length of the spin as $S$, the concrete expressions of $h_{1}(\bm{k}),h_{2}(\bm{k}),\Delta_{1}(\bm{k}),$ and $\Delta_{2}(\bm{k})$ are given by
\begin{widetext}
\begin{align}
&h_{1}(\bm{k})=
S\left(
\begin{array}{ccc}
8J+J'+2\kappa_{A} &-J\gamma_{AB}(\bm{k})+{\rm i}D\delta_{AB}(\bm{k})  &-J\gamma_{AC}(\bm{k})+{\rm i}D\delta_{AC}(\bm{k})   \\
-J\gamma_{AB}^{*}(\bm{k})-{\rm i}D\delta_{AB}^{*}(\bm{k})  &8J+J'+2\kappa_{B} &-J\gamma_{BC}(\bm{k})+{\rm i}D\delta_{BC}(\bm{k})   \\
-J\gamma_{AC}^{*}(\bm{k})-{\rm i}D\delta_{AC}^{*}(\bm{k})&-J\gamma_{BC}^{*}(\bm{k})-{\rm i}D\delta_{BC}^{*}(\bm{k}) &8J+J'+2\kappa_{C}   \\
\end{array} 
\right), \\
&h_{2}(\bm{k})=
S\left(
\begin{array}{ccc}
0 &-J_{-}\gamma_{AB}(\bm{k})  &-J_{-}\gamma_{AC}(\bm{k})    \\
J_{-}\gamma_{AB}^{*}(\bm{k}) &0  &-J_{-}\gamma_{BC}(\bm{k})    \\
J_{-}\gamma_{AC}^{*}(\bm{k})  &J_{-}\gamma_{BC}^{*}(\bm{k})   &0    \\
\end{array} 
\right), \\
&\Delta_{1}(\bm{k})=
S\left(
\begin{array}{ccc}
J' &J_{+}\gamma_{AB}(\bm{k})  &J_{+}\gamma_{AC}(\bm{k})    \\
J_{+}\gamma_{AB}^{*}(\bm{k}) &J'  &J_{+}\gamma_{BC}(\bm{k})    \\
J_{+}\gamma_{AC}^{*}(\bm{k})  &J_{+}\gamma_{BC}^{*}(\bm{k})   &J'    \\
\end{array} 
\right), \\
&\Delta_{2}(\bm{k})=0,
\end{align}
\end{widetext}
where
\begin{align}
&\gamma_{AB}(\bm{k})=1+e^{{\rm i} k_{1}}+e^{-{\rm i} k_{2}}+e^{{\rm i}k_{12}}, \\
&\gamma_{AC}(\bm{k})=1+e^{{\rm i} k_{1}}+e^{-{\rm i} k_{3}}+e^{{\rm i}k_{13}}, \\
&\gamma_{BC}(\bm{k})=1+e^{{\rm i} k_{2}}+e^{-{\rm i} k_{3}}+e^{{\rm i}k_{23}}, \\
&\delta_{AB}(\bm{k})=-s_{3}\left(1-e^{{\rm i} k_{1}}-e^{-{\rm i} k_{2}}+e^{{\rm i}k_{12}}\right) ,\\
&\delta_{AC}(\bm{k})=s_{2}\left(1-e^{{\rm i} k_{1}}-e^{-{\rm i} k_{3}}+e^{{\rm i}k_{13}}\right), \\
&\delta_{BC}(\bm{k})=-s_{1}\left(1-e^{{\rm i} k_{2}}-e^{-{\rm i} k_{3}}+e^{{\rm i}k_{23}}\right).
\end{align}
Here $J_{\pm}$ is defined as $J_{\pm}=(J_{x}\pm J_{y})/2$.

\begin{figure}[H]
\centering
  \includegraphics[width=8.5cm]{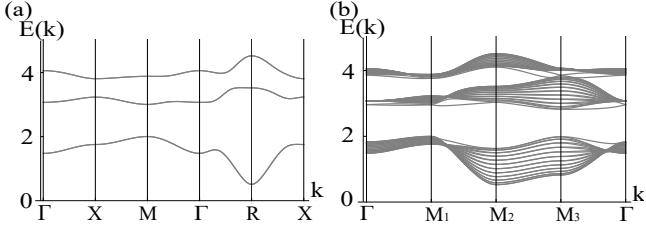}
\caption{(a) Band structure of a slab with (001) face. Parameters are chosen to be $JS=0.03,DS=0.4,J_{+}S=0.1,J_{-}S=0.1,J'S=1.0,\kappa_{A}S=1.35,\kappa_{B}S=0.5,\kappa_{C}S=1.0,s_{1}=-(2+\sqrt{2})/4,s_{2}=(2-\sqrt{2})/4,s_{3}=1/2$. 
(b) Band structure for a slab with (001) face. 
}
\label{fig:band_pe}
\end{figure}

Using the numerical implementation described in Ref.~\cite{Fukui07}, we calculate the topological invariants of the system.
The results are shown in Table~\ref{table:index_pe}. 
We note that the realization of nontrivial topological phases depend on the direction of ordered spins $\bm{s}$. 
For example,  in the case of $\bm{s}=(0,0,1)$, the band structure is topologically trivial, while the index $\nu_{0}^{n}=1$ $(n=1,2)$ in Table~\ref{table:index_pe} implies that the system is in the strong topological phase if $s_{1}=-(2+\sqrt{2})/4,s_{2}=(2-\sqrt{2})/4,s_{3}=1/2$.
In order to see the correspondence between the topological invariants in Table~\ref{table:index_pe} and the number of Dirac cones, let us plot the band structure of the system.
The band structure of the bulk system and a slab with (001) face are shown in Fig.~\ref{fig:band_pe}.
As far as we investigate this system, the strong topological phase is realized only in the system with the indirect gap.

\begin{table}[H]
\caption{The topological invariants of the perovskite lattice system. Parameters are chosen to be $JS=0.03,DS=0.4,J_{+}S=0.1,J_{-}S=0.1,J'S=1.0,\kappa_{A}S=1.35,\kappa_{B}S=0.5,\kappa_{C}S=1.0,s_{1}=-(2+\sqrt{2})/4,s_{2}=(2-\sqrt{2})/4,s_{3}=1/2$. The index $n=1,2,3$ denotes the band with $n$th highest energy in the particle space. 
}
\begin{center}
{\tabcolsep=2mm
  \begin{tabular}{cccccccc}
    \hline
    $n$  & $\nu_{x,0}^{n}$ & $\nu_{x,\pi}^{n}$ &$\nu_{y,0}^{n}$  & $\nu_{y,\pi}^{n}$ & $\nu_{z,0}^{n}$ & $\nu_{z,\pi}^{n}$ & $(\nu_{0}^{n};\nu_{x}^{n},\nu_{y}^{n},\nu_{z}^{n})$  \\ \hline
    1  & 1 & 0& 0 & 1& 0 & 1 & (1;0,1,1)  \\ \hline
    2  & 1 & 0& 0 & 1& 0 & 1 & (1;0,1,1)  \\ \hline
    3  & 0 & 0& 0 & 0& 0 & 0 & (0;0,0,0)  \\ \hline
  \end{tabular}
}
\end{center}
\label{table:index_pe}
\end{table}
\noindent
\noindent
As shown in Fig.~\ref{fig:band_pe}(b), a single Dirac cone exists at the ${\rm M}_{3}$ point of the upper gap.
This corresponds to the summation of the strong indices of the lower and middle bands $\nu_{0}^{2}+\nu_{0}^{3}=1$.
This Dirac cone is robust against disorder which does not break the pseudo-time-reversal symmetry.
The topological invariants of the lower band are all zero.
This indicates the absence of the Dirac cones in the gap between the lower and middle bands.

\begin{center}
\bf{Appendix E: Third model of 3D topological magnon systems: pyrochlore lattice system}
\end{center}

\begin{figure}[H]
\centering
  \includegraphics[width=7cm]{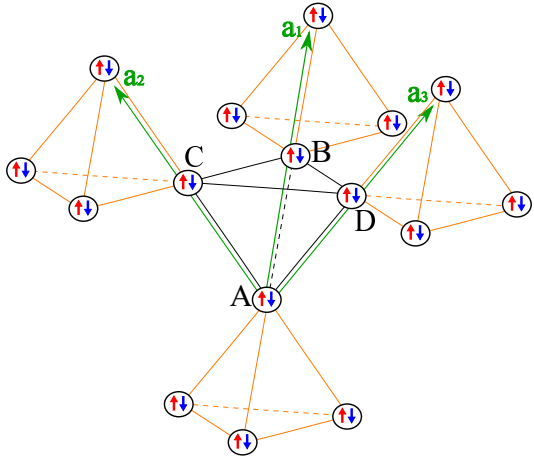}
\caption{The pyrochlore lattice system. Up and down spins are localized at each site. Four sublattices are denoted as $A,B,C$, and $D$. The vectors $\bm{a}_{i}$ $(i=1,2,3)$ are the lattice basis vectors.
}\label{fig:cell_py}
\end{figure}

We give another model of 3D magnon topological systems.
This is the system in which the same number of up and down spins are localized at the same site of the pyrochlore lattice.
The schematic picture of the system is shown in Fig.~\ref{fig:cell_py}.
\noindent
The Hamiltonian of the system is written as
\begin{align}
H=H_{\rm DM}+H_{J}+H_{\rm XY}+H_{J'}+H_{\kappa}, \label{eq:Ham_py}
\end{align}
where $H_{\rm DM}$ and $H_{J}$ are DM and ferromagnetic interactions between nearest neighbor spins which point in the same direction.
The third term $H_{\rm XY}$ is the anisotropic XY interaction between nearest neighbor spins pointing in opposite directions.
The details of the DM interactions in the pyrochlore lattice are shown in Appendix F.
The other terms $H_{J'}$ and $H_{\kappa}$ are the antiferromagnetic interaction between two spins on the same site and the single ion anisotropy, respectively.
The single ion anisotropy opens the gap of the band structure.
The Hamiltonians $H_{\rm DM},H_{J},H_{J'}$, and $H_{\kappa}$ are given by
\begin{align}
&H_{\rm DM}=\sum_{\langle i,j\rangle,s={\rm u,d}} \bm{D}_{ij}\cdot (\bm{S}_{i,s}\times\bm{S}_{j,s}) ,\\
&H_{J}=-J\sum_{\langle i,j\rangle,s={\rm u,d}}\bm{S}_{i,s}\cdot\bm{S}_{j,s},  \\
&H_{J'}=J'\sum_{i}\bm{S}_{i,{\rm u}}\cdot\bm{S}_{i,{\rm d}} , \\
&H_{\kappa}=-\sum_{X=A,B,C,D}\sum_{i\in X,s={\rm u,d}}\kappa_{X}(S_{i,s}^{z})^{2},
\end{align}
where operators $\bm{S}_{i,{\rm u}}$ and $\bm{S}_{i,{\rm d}}$ denote up and down spins on site $i$, respectively.
The anisotropic XY interaction $H_{\rm XY}$ is written as
\begin{widetext}
\begin{align}
H_{\rm XY} 
&=\sum_{\bm{R}}\bm{S}_{\rm u}(\bm{R},A)J_{\rm XY}\left[\bm{S}_{\rm d}(\bm{R},B)\!+\!\bm{S}_{\rm d}(\bm{R}\!-\!\bm{a}_{1},B)\right]\!+\!\bm{S}_{\rm u}(\bm{R},A)J_{\rm XY}\left[\bm{S}_{\rm d}(\bm{R},C)\!+\!\bm{S}_{\rm d}(\bm{R}\!-\!\bm{a}_{2},C)\right] \nonumber \\
&\hspace{7mm}\!+\!\bm{S}_{\rm u}(\bm{R},A)J_{\rm XY}\left[\bm{S}_{\rm d}(\bm{R},D)\!+\!\bm{S}_{\rm d}(\bm{R}\!-\!\bm{a}_{3},D)\right]\!+\!\bm{S}_{\rm u}(\bm{R},B)J_{\rm XY}\left[\bm{S}_{\rm d}(\bm{R},C)\!+\!\bm{S}_{\rm d}(\bm{R}\!-\!\bm{a}_{21},C)\right] \nonumber \\
&\hspace{7mm}\!+\!\bm{S}_{\rm u}(\bm{R},B)J_{\rm XY}\left[\bm{S}_{\rm d}(\bm{R},D)\!+\!\bm{S}_{\rm d}(\bm{R}\!-\!\bm{a}_{31},D)\right]\!+\!\bm{S}_{\rm u}(\bm{R},C)J_{\rm XY}\left[\bm{S}_{\rm d}(\bm{R},D)\!+\!\bm{S}_{\rm d}(\bm{R}\!-\!\bm{a}_{32},D)\right] \nonumber \\
&\hspace{7mm}\!+\!\bm{S}_{\rm d}(\bm{R},A)J_{\rm YX}\left[\bm{S}_{\rm u}(\bm{R},B)\!+\!\bm{S}_{\rm u}(\bm{R}\!-\!\bm{a}_{1},B)\right]\!+\!\bm{S}_{\rm d}(\bm{R},A)J_{\rm YX}\left[\bm{S}_{\rm u}(\bm{R},C)\!+\!\bm{S}_{\rm u}(\bm{R}\!-\!\bm{a}_{2},C)\right] \nonumber \\
&\hspace{7mm}\!+\!\bm{S}_{\rm d}(\bm{R},A)J_{\rm YX}\left[\bm{S}_{\rm u}(\bm{R},D)\!+\!\bm{S}_{\rm u}(\bm{R}\!-\!\bm{a}_{3},D)\right]\!+\!\bm{S}_{\rm d}(\bm{R},B)J_{\rm YX}\left[\bm{S}_{\rm u}(\bm{R},C)\!+\!\bm{S}_{\rm u}(\bm{R}\!-\!\bm{a}_{21},C)\right] \nonumber \\
&\hspace{7mm}\!+\!\bm{S}_{\rm d}(\bm{R},B)J_{\rm YX}\left[\bm{S}_{\rm u}(\bm{R},D)\!+\!\bm{S}_{\rm u}(\bm{R}\!-\!\bm{a}_{31},D)\right]\!+\!\bm{S}_{\rm d}(\bm{R},C)J_{\rm YX}\left[\bm{S}_{\rm u}(\bm{R},D)\!+\!\bm{S}_{\rm u}(\bm{R}\!-\!\bm{a}_{32},D)\right].\nonumber 
\end{align}
\end{widetext}

Applying the Holstein-Primakoff transformation, we rewrite the Hamiltonian (\ref{eq:Ham_py}) in the same form as
Eq.~(\ref{eq:Ham_gen}).
We take the operators $\bm{b}_{\uparrow}^{\dagger}(\bm{k})$ and $\bm{b}_{\downarrow}^{\dagger}(\bm{k})$ in Eq.~(\ref{eq:opvec}) as
\begin{align}
&\bm{b}_{\uparrow}^{\dagger}(\bm{k})=[b_{\rm u}^{\dagger}(\bm{k},A),b_{\rm u}^{\dagger}(\bm{k},B),b_{\rm u}^{\dagger}(\bm{k},C),b_{\rm u}^{\dagger}(\bm{k},D)],  \\
&\bm{b}_{\downarrow}^{\dagger}(\bm{k})=[b_{\rm d}^{\dagger}(\bm{k},A),b_{\rm d}^{\dagger}(\bm{k},B),b_{\rm d}^{\dagger}(\bm{k},C),b_{\rm d}^{\dagger}(\bm{k},D)] .
\end{align}
The concrete expressions of $h_{1}(\bm{k}),h_{2}(\bm{k}),\Delta_{1}(\bm{k})$, and $\Delta_{2}(\bm{k})$ are given by
\begin{widetext}
\begin{align}
&h_{1}(\bm{k})=S\left(
\begin{array}{cccc}
6J+J'+2\kappa_{A} &-(J-{\rm i}\tilde{D})\gamma_{1}^{*}(\bm{k}) &-(J+{\rm i}\tilde{D})\gamma_{2}^{*}(\bm{k}) &-J\gamma_{3}^{*}(\bm{k}) \\
-(J+{\rm i}\tilde{D})\gamma_{1}(\bm{k}) &6J+J'+2\kappa_{B} &-(J+2{\rm i}\tilde{D})\gamma_{21}^{*}(\bm{k}) &-(J-3{\rm i}\tilde{D})\gamma_{31}^{*}(\bm{k}) \\
-(J-{\rm i}\tilde{D})\gamma_{2}(\bm{k}) &-(J-2{\rm i}\tilde{D})\gamma_{21}(\bm{k}) &6J+J'+2\kappa_{C} &-(J+3{\rm i}\tilde{D})\gamma_{32}^{*}(\bm{k})  \\
-J\gamma_{3}(\bm{k}) &-(J+3{\rm i}\tilde{D})\gamma_{31}(\bm{k}) &-(J-3{\rm i}\tilde{D})\gamma_{32}(\bm{k}) &6J+J'+2\kappa_{D}  
\end{array}
\!\right)\!,  \\
&h_{2}(\bm{k})=S\left(
\begin{array}{cccc}
0 &-J_{-}\gamma_{1}^{*}(\bm{k}) &-J_{-}\gamma_{2}^{*}(\bm{k}) &-J_{-}\gamma_{3}^{*}(\bm{k}) \\
J_{-}\gamma_{1}(\bm{k}) &0 &-J_{-}\gamma_{21}^{*}(\bm{k}) &-J_{-}\gamma_{31}^{*}(\bm{k}) \\
J_{-}\gamma_{2}(\bm{k}) &J_{-}\gamma_{21}(\bm{k}) &0 &-J_{-}\gamma_{32}^{*}(\bm{k})  \\
J_{-}\gamma_{3}(\bm{k}) &J_{-}\gamma_{31}(\bm{k}) &J_{-}\gamma_{32}(\bm{k}) &0  
\end{array}
\right),  \\
&\Delta_{1}(\bm{k})=S\left(
\begin{array}{cccc}
J' &J_{+}\gamma_{1}^{*}(\bm{k}) &J_{+}\gamma_{2}^{*}(\bm{k}) &J_{+}\gamma_{3}^{*}(\bm{k}) \\
J_{+}\gamma_{1}(\bm{k}) &J' &J_{+}\gamma_{21}^{*}(\bm{k}) &J_{+}\gamma_{31}^{*}(\bm{k}) \\
J_{+}\gamma_{2}(\bm{k}) &J_{+}\gamma_{21}(\bm{k}) &J' &J_{+}\gamma_{32}^{*}(\bm{k})  \\
J_{+}\gamma_{3}(\bm{k}) &J_{+}\gamma_{31}(\bm{k}) &J_{+}\gamma_{32}(\bm{k}) &J'  
\end{array}
\right),  \\
&\Delta_{2}(\bm{k})=0,
\end{align}
\end{widetext}
where $\gamma_{i}(\bm{k})=1+e^{{\rm i} k_{i}},\gamma_{ij}(\bm{k})=1+e^{{\rm i}k_{ij}}$, and $\tilde{D}=\frac{\sqrt{3}}{6}D$.

\begin{figure}[H]
\centering
  \includegraphics[width=8.5cm]{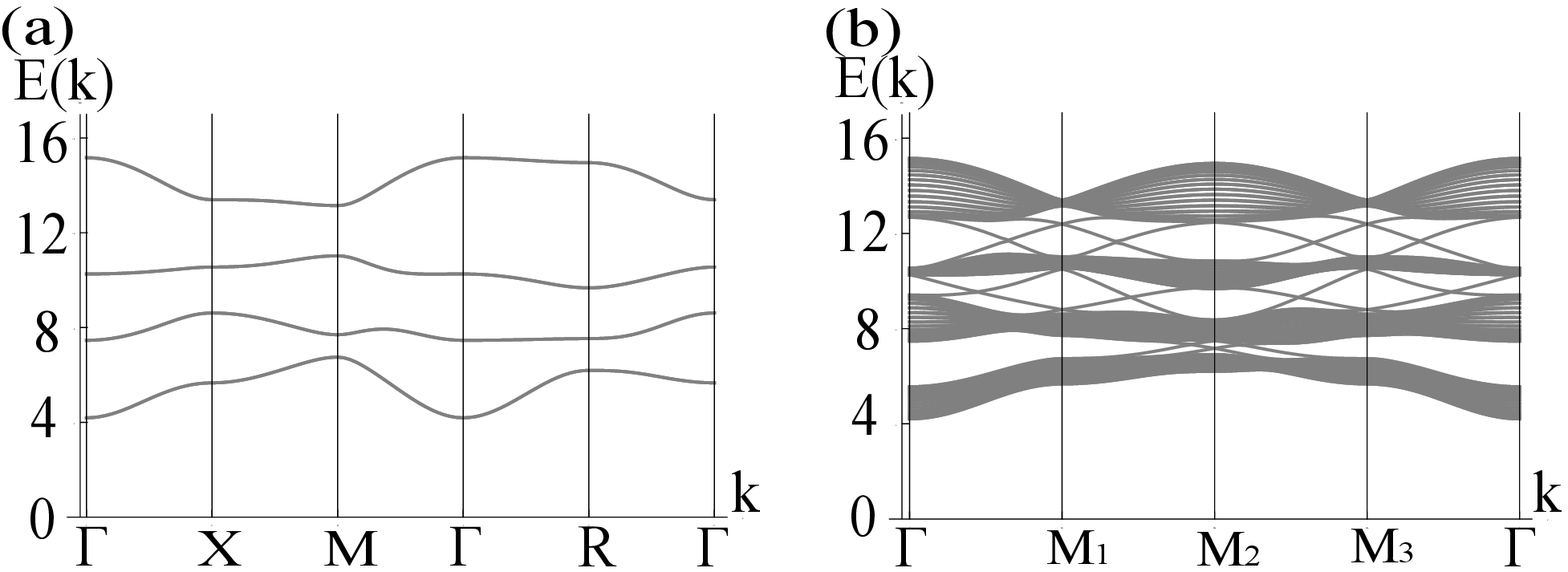}
\caption{(a) Bulk band structure of the pyrochlore lattice system. Taking $\bm{a}_{1}=(1,0,0),\bm{a}_{2}=(0,1,0),\bm{a}_{3}=(0,0,1)$ we here deform the pyrochlore lattice into an equivalent cubic lattice.
(b) The band structure for a slab with (001) face. In both cases, parameters are chosen to be $JS=0.1,\tilde{D}S=0.4,J_{+}S=0.8,J_{-}S=0.6,J'S=1.0,\kappa_{A}S=3.0,\kappa_{B,C,D}S=4.5$. 
}\label{fig:band_py}
\end{figure}

Using the numerical implementation described in Ref.~\cite{Fukui07}, we calculate the topological invariants of the system.
The results are shown in Table~\ref{table:index_py}.
Table~\ref{table:index_py} suggests that the system has even, even, and odd number of Dirac cones in the highest, second highest, and the lowest band gaps, respectively. 
In order to confirm this, we calculate the band structure of the system.
The band structure of the bulk system and a slab with (001) face is shown in Fig.~\ref{fig:band_py}.
A single Dirac cone exists at the ${\rm M_{2}}$ point in the lowest band gap.
This corresponds to the strong topological index $\nu_{0}^{4}=1$.
The second highest and the lowest band gaps have two Dirac cones. 
This means that the summations of the strong topological index over the lowest two and three bands are zero:
\begin{align}
&\nu_{0}^{3}+\nu_{0}^{4}=0 \hspace{3mm}{\rm mod}\hspace{1mm}2, \\
&\nu_{0}^{2}+\nu_{0}^{3}+\nu_{0}^{4}=0 \hspace{3mm}{\rm mod}\hspace{1mm}2.
\end{align}

\begin{table}[H]
\caption{The topological invariants of the pyrochlore lattice system. Parameters are chosen to be $JS=0.1,\tilde{D}S=0.4,J_{+}S=0.8,J_{-}S=0.6,J'S=1.0,\kappa_{A}S=3.0,\kappa_{B,C,D}S=4.5$. The index $n=1,2,3,4$ denotes the band with the $n$th highest energy in the particle space.
}
\begin{center}
{\tabcolsep=1mm
  \begin{tabular}{cccccccc}
    \hline
    $n$  & $\nu_{x,0}^{n}$ & $\nu_{x,\pi}^{n}$ &$\nu_{y,0}^{n}$  & $\nu_{y,\pi}^{n}$ & $\nu_{z,0}^{n}$ & $\nu_{z,\pi}^{n}$ & $(\nu_{0}^{n};\nu_{x}^{n},\nu_{y}^{n},\nu_{z}^{n})$  \\ \hline
    1  & 1 & 1& 1 & 1& 1 & 1 & (0;1,1,1)  \\ \hline
    2  & 0 & 0& 0 & 0& 0 & 0 & (0;0,0,0)  \\ \hline
    3  & 1 & 0& 1 & 0& 1 & 0 & (1;0,0,0)  \\ \hline
    4  & 0 & 1& 0 & 1& 0 & 1 & (1;1,1,1)  \\ \hline
  \end{tabular}
}
\end{center}
\label{table:index_py}
\end{table}
\noindent
\noindent

\begin{figure}[H]
\centering
  \includegraphics[width=5cm]{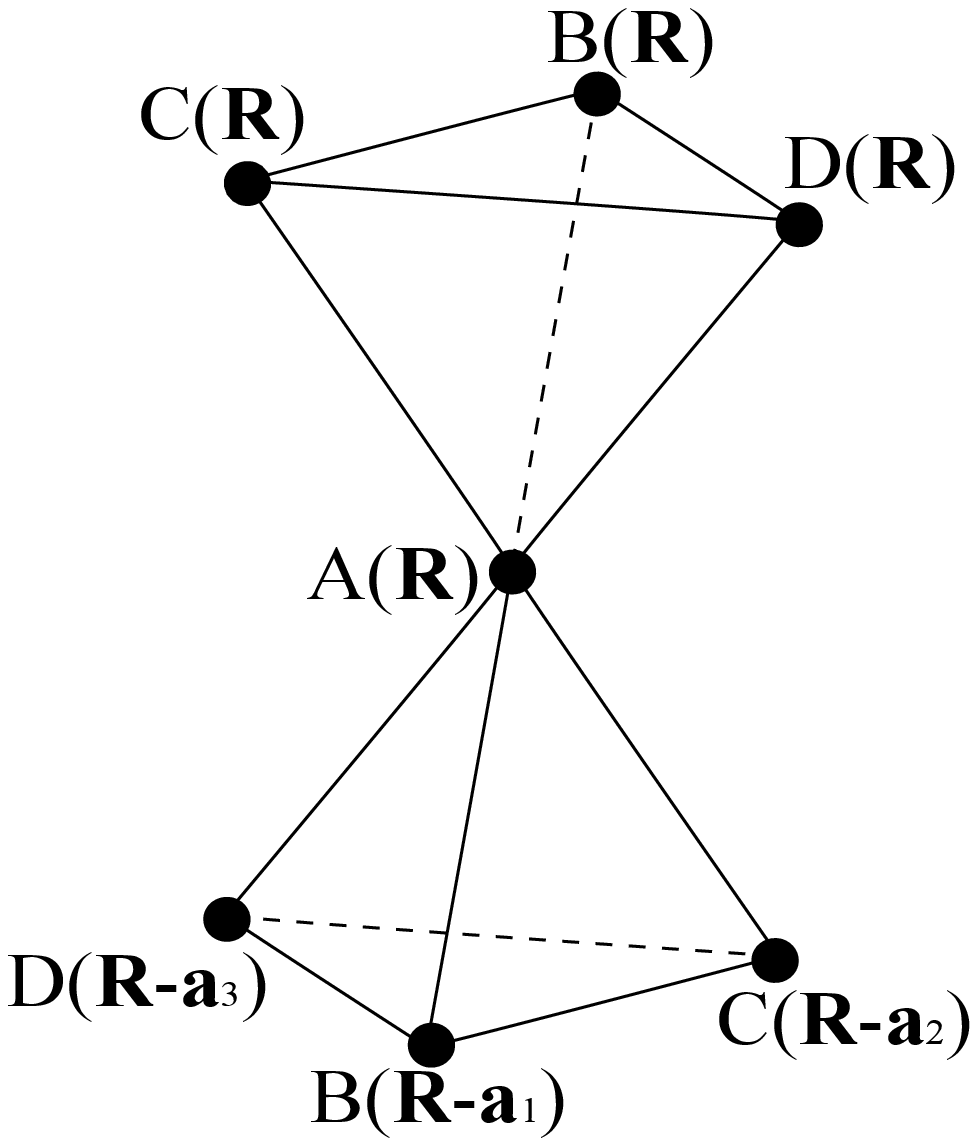}
\caption{The nearest neighbor bonds of the pyrochlore lattice.
}\label{fig:DM_py}
\end{figure}

\begin{center}
\bf{Appendix F: Details of the DM interaction in the pyrochlore lattice system}
\end{center}
Here we write the details of DM vectors in the pyrochlore lattice system.
The nearest neighbor bonds of the pyrochlore lattice are shown in Fig.~\ref{fig:DM_py}. 
Here $X(\bm{R})$ is the $X$ sublattice in the unit cell with the lattice vector $\bm{R}$.
\noindent
We write the DM vector between two sites $X(\bm{R})$ and $X'(\bm{R'})$ as $\bm{D}(X(\bm{R}),X'(\bm{R'}))$($\equiv \bm{D}_{ij}$).
The unit vector from the site $X(\bm{R})$ to $X'(\bm{R})$ is defined as $\overrightarrow{XX'}$.
Each DM vector of the pyrochlore lattice is written as
\begin{align}
&\bm{D}(A(\bm{R}),B(\bm{R}))=\bm{D}(A(\bm{R}),B(\bm{R}-\bm{a}_{1}))=D\overrightarrow{CD}, \\
&\bm{D}(A(\bm{R}),C(\bm{R}))=\bm{D}(A(\bm{R}),C(\bm{R}-\bm{a}_{2}))=D\overrightarrow{DB}, \\
&\bm{D}(A(\bm{R}),D(\bm{R}))=\bm{D}(A(\bm{R}),D(\bm{R}-\bm{a}_{3}))=D\overrightarrow{BC}, \\
&\bm{D}(B(\bm{R}),C(\bm{R}))=\bm{D}(B(\bm{R}-\bm{a}_{1}),C(\bm{R}-\bm{a}_{2}))=D\overrightarrow{AD}, \\
&\bm{D}(B(\bm{R}),D(\bm{R}))=\bm{D}(B(\bm{R}-\bm{a}_{1}),D(\bm{R}-\bm{a}_{3}))=D\overrightarrow{CA}, \\
&\bm{D}(C(\bm{R}),D(\bm{R}))=\bm{D}(C(\bm{R}-\bm{a}_{2}),D(\bm{R}-\bm{a}_{3}))=D\overrightarrow{AB}. 
\end{align}


\begin{thebibliography}{99}
\bibitem{Energy_analytical} As shown in Appendix C, the energy eigenvalues can be obtained analytically when $\Gamma=0$.
\end{thebibliography}
\end{document}